\documentclass[fleqn,usenatbib]{mnras}

\usepackage{newtxtext}%,newtxmath}
\usepackage[T1]{fontenc}
\usepackage{ae,aecompl}
\usepackage{booktabs}

\usepackage{graphicx}	% Including figure files
\usepackage{amsmath}	% Advanced maths commands
\usepackage{amssymb}	% Extra maths symbols
\usepackage{natbib}
\bibpunct{(}{)}{;}{a}{}{,} 
\usepackage{xcolor}
\newcommand{\es}[1]{}
\usepackage[caption=false]{subfig}
\usepackage{pifont}

\newcommand{\highlight}[1]{}

\usepackage{txfonts}
\title[\texttt{saprEMo}]{saprEMo: a simplified algorithm for predicting detections of electromagnetic transients in surveys}

\author[S. Vinciguerra et al.]{
			S. Vinciguerra$^{1,2}$, 
          M. Branchesi$^{3,4}$,
          R. Ciolfi$^{5,6}$,
          I. Mandel$^{1,7}$,
          C. J. Neijssel$^{1}$,
          \and
          G. Stratta $^{8,9}$
\\
$^{1}$ Institute of Gravitational Wave Astronomy and School of Physics and Astronomy, University of Birmingham,  B15 2TT, Birmingham, UK\\
$^{2}$ Max Planck Institute for Gravitational Physics (Albert Einstein Institute), D-30167 Hannover, Germany\\
$^{3}$ Gran Sasso Science Institute, Viale Francesco Crispi, 7, 67100 L'Aquila AQ, Italy  \\
$^{4}$ INFN, Laboratori Nazionali del Gran Sasso, I-67100 Assergi, Italy \\
$^{5}$ INAF, Osservatorio Astronomico di Padova,
Vicolo dell'Osservatorio 5, I-35122 Padova, Italy\\
$^{6}$ INFN-TIFPA, Trento Institute for Fundamental Physics and Applications, Via Sommarive 14, I-38123 Trento, Italy\\
$^{7}$ Monash Centre for Astrophysics, School of Physics and Astronomy, Monash University, Clayton, Victoria 3800, Australia\\
$^{8}$  University of Urbino "Carlo Bo", Dipartimento di Scienze di Base e Fondamenti/Physics section, Via Santa Chiara, 27, 61029 Urbino, Italy\\
$^{9}$ INFN, Sezione di Firenze, I-50019 Sesto Fiorentino, Firenze, Italy
}

\date{Accepted 14 Dec 2018. Received 23 Sep 2018; in original form 23 Sep 2018}

\pubyear{2018}
\begin{document}
\label{firstpage}
\pagerange{\pageref{firstpage}--\pageref{lastpage}}
\maketitle

% Abstract of the paper
\begin{abstract}
The multi-wavelength detection of GW170817 has inaugurated multi-messenger astronomy. The next step consists in interpreting observations coming from population of gravitational wave sources. We introduce saprEMo, a tool aimed at predicting the number of electromagnetic signals characterised by a specific light curve and spectrum, expected in a particular sky survey. By looking at past surveys, saprEMo allows us to constrain models of electromagnetic emission or event rates. Applying saprEMo to proposed astronomical missions/observing campaigns provides a perspective on their scientific impact and tests the effect of adopting different observational strategies. For our first case study, we adopt a model of spindown-powered X-ray emission predicted for a binary neutron star merger producing a long-lived neutron star. We apply saprEMo on data collected by {\it XMM-Newton}  and {\it Chandra} and during $10^4$s of observations with the mission concept {\it THESEUS}. We demonstrate that our emission model and binary neutron star merger rate imply the presence of some signals in the {\it XMM-Newton} catalogs. We also show that the new class of X-ray transients found by Bauer et al. in the {\it Chandra} Deep Field-South is marginally consistent with the expected rate. Finally, by studying the mission concept {\it THESEUS}, we demonstrate the substantial impact of a much larger field of view in searches of X-ray transients.
\end{abstract}

\begin{keywords}
EM follow-up models --
BNS mergers --
rates -- 
short gamma-ray bursts
\end{keywords}

%%%%%%%%%%%%%%%%%%%%%%%%%%%%%%%%%%%%%%%%%%%%%%%%%%

%%%%%%%%%%%%%%%%% BODY OF PAPER %%%%%%%%%%%%%%%%%%

\section{Introduction}
GW170817 (\citealt{GW170817}) has just opened the era of multi-messenger astronomy (\citealt{MMA}).
The first coincident set of gravitational-waves and electromagnetic observations has already provided an extraordinary insight into 
the physics of the binary neutron star mergers.
Among the key results of this revolutionary discovery is the confirmation of the association between the merger of two neutron stars (NSs) and (at least some) short gamma ray bursts (SGRBs) (\citealt{MMA} and refs. therein). 
The last radio VLBI observations demonstrate 
that a narrow jet was formed
and prove the association with a classical SGRB (see 
\citealt{troja2017x, mooley2018mildly, margutti2018binary, d2018evolution, lyman2018optical, dobie2018turnover,mooley2018superluminal,ghirlanda2018re}).

The intense multi-wavelength follow-ups of gamma ray bursts in the last decade have revealed
new and unexpected features, such as early and late X-ray flares, 
extended emission, 
and X-ray plateaus (e.g., \citealt{berger2014short} and refs. therein).
The challenges posed by this rich astronomical scenario motivated
a growing interest of the community in investigating compact binary mergers from both the theoretical and observational points of view.
Intensified theoretical efforts have been 
dedicated to explain these observations and coherently explore these and other possible electromagnetic signals generated by these sources.

In order to validate the variety of proposed theoretical scenarios in the context of multi-messenger astronomy with compact binary mergers, we present \texttt{saprEMo}.

We developed \texttt{saprEMo}, 
a {\it Simplified Algorithm for PRedicting ElectroMagnetic Observations}, to evaluate how many electromagnetic (EM) signals, characterised by a specific light curve and spectrum, should be present in a data set
given some overall characteristics of the astronomical survey and a cosmological rate of compact binary mergers.
Predictions can be used both to validate the theoretical scenarios against already collected data and to critically examine the scientific means of future missions and their observational strategies. While we use compact binary mergers as the prime multi-messenger targets, \texttt{saprEMo} can also be applied to other type of transients (e.g. core-collapse supernovae).

We describe the main features of \texttt{saprEMo} in Section \ref{S2:saprEMo_outline}. 
As first case study, we use \texttt{saprEMo} to investigate the X-ray emission 
from Binary NS (BNS) mergers leading to the formation of a long-lived and strongly magnetized NS, following the model of \citealt{siegel2016electromagnetic,siegel2016electromagneticII} (see Section \ref{S3_1:Xray_model}).
We apply \texttt{saprEMo} to present X-ray surveys, collected by {\it XMM-Newton} \citep{jansen2001xmm, struder2001european, turner2001european}
and {\it Chandra} \citep{weisskopf1996advanced, weisskopf2000chandra}, and study the prospectives of the mission concept {\it THESEUS} \citep{Amati2017}. Results are reported in Section \ref{S3_3:Xray_results} and discussed in Section \ref{S4:Discussion}. Finally, in Section \ref{S5:Conclusions} we draw our conclusions summarising our first results and outlining the main features and primary scopes of \texttt{saprEMo}.
Throughout the paper we assume a flat cosmology with: $H_0= 70\mathrm{~km~s^{-1}~Mpc^{-1}}$, $\Omega_{M} = 0.3$ and $\Omega_{\Lambda} = 0.7$.
%%%%%%%%%%%%%%%%%%%%%%%%%%%%%%%%%%%%%%%%%%%%%%%%%%%
\section{saprEMo outline}
\label{S2:saprEMo_outline}
%%%%%%%%%%%%%%%%%%%%%%%%%%%%%%%%%%%%%%%%%%%%%%%%%%%
\texttt{saprEMo} is a Python algorithm designed to 
predict how many detectable electromagnetic signals, associated with a specific EM emission ({\it EMe}) model, are present in a survey {\it S}. 
The full code and a short manual are publicly available at \texttt{https://github.com/saprEMo/source\_code}. \\
According to instrument limitations (such field of view and spectral sensitivity), \texttt{saprEMo} estimates the number of signals
whose emission flux $F$ at the observer is above the flux limit $F_{lim}$ of the {\it S} survey.
Accounting for the energy dependency of the survey sensitivity, we define detections
on a instantaneous flux-based criterion: $\exists~ g, t' ~\mid F_g\left(t'\right)>F_{lim,g}$, where $g$ labels the spectral band of the survey. We therefore simplify our analysis by treating
detectability in each band independently, i.e., a source is considered to be detected if and only if it can be independently detected in at least one instrument band.  More realistic treatments include flux integration over the observation time and noise simulation (see \citealt{carbone2016calculating} and references therein for a discussion of these data analysis aspects).

\texttt{saprEMo} does not directly consider the actual sky locations observed by the survey {\it S} (even when applied to archival data) and instead focuses on accounting for cosmological distances, relying on the isotropy of the Universe.
%%%%%%%%%%%%%%%%%%%%%%%%%%%%%%%%%%%%%%%%%%%%%%%%%%%
\subsection{Core analysis}
\begin{figure}
   \centering
		\begin{minipage}[b]{0.5\textwidth}
        	\centering
        	\includegraphics[width=8.cm]{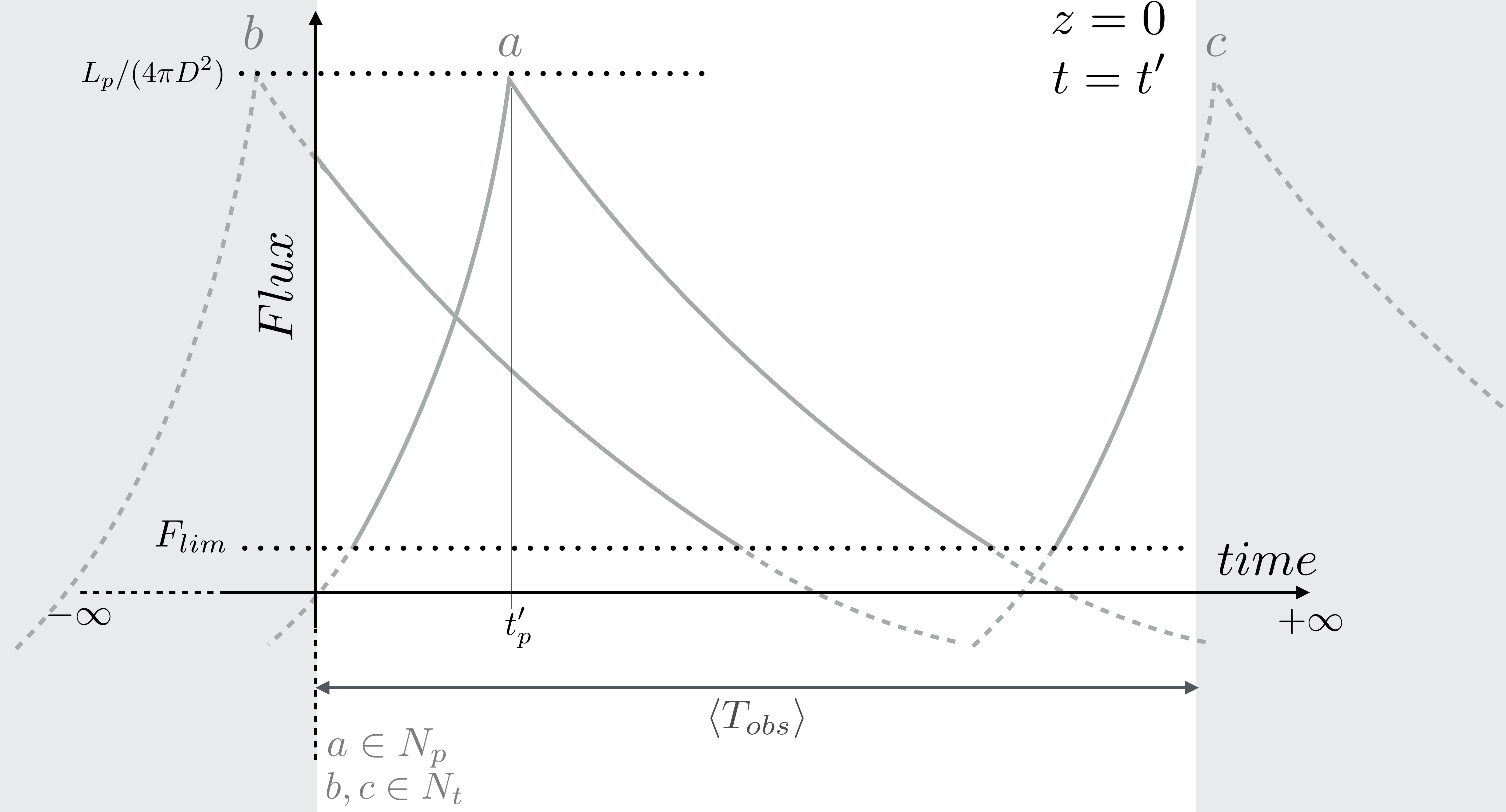}
   			\caption{Schematic representation of a peak ($t_p \in \left<T_{obs}\right>$)  signal ($a$) and tails ($b,~c$). The solid curves represent the part of signals {\it EMe} visible during the exposure time at the observer, the dashed components are the missed (because of time or flux restrictions) part of the emissions. The upper dotted line shows the peak flux $F_p = L_p/(4\pi D^2)$, the lower line the flux limit of the survey.}
            \label{fig:1}
		\end{minipage}
        \hfill
		\begin{minipage}[b]{0.5\textwidth}
        \centering
        \includegraphics[width=8.cm]{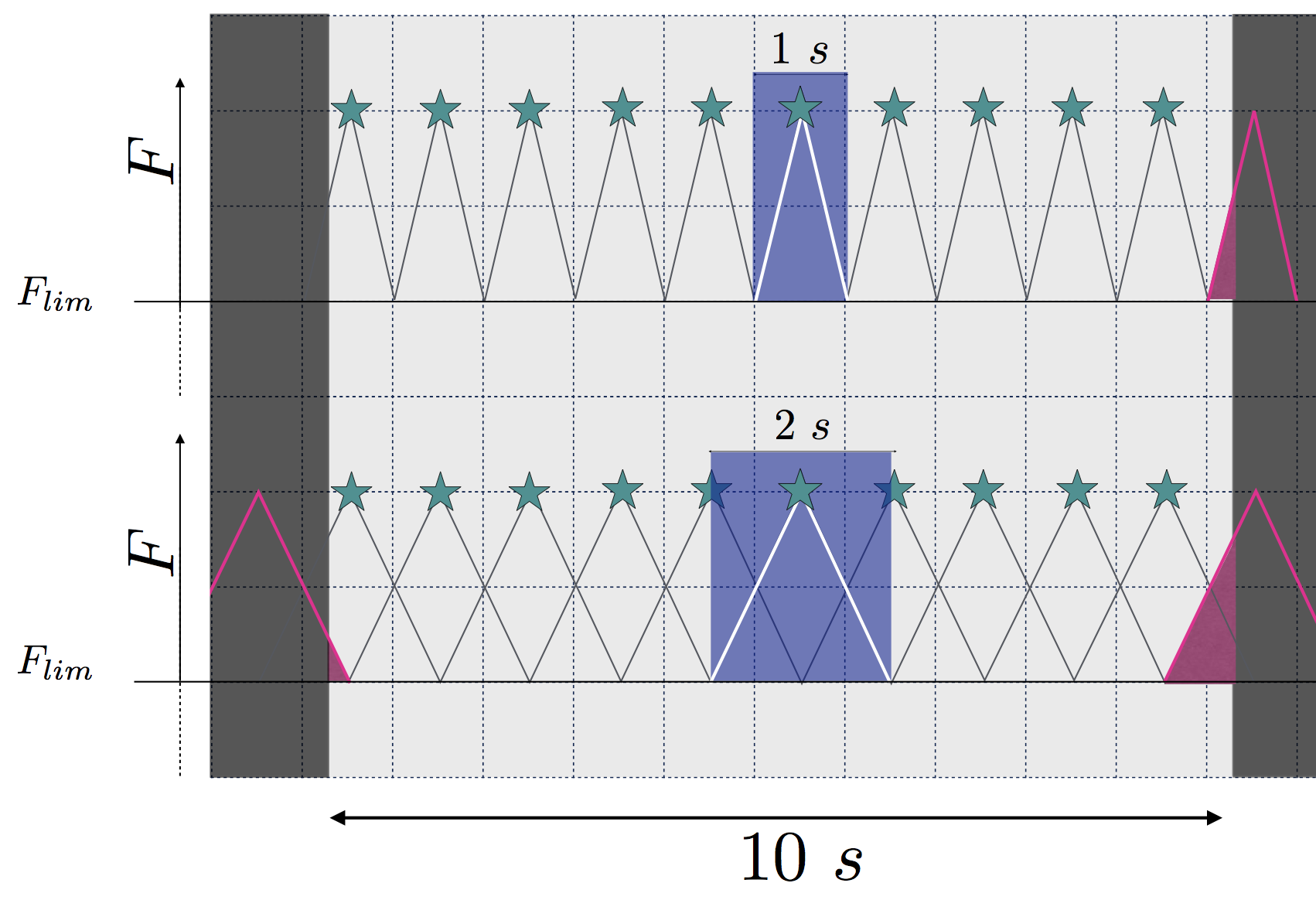}
  		  			\caption{Simplified examples of peaks and tails. In both upper and lower panels we report the detected flux (in arbitrary units on the y-axis) as a faction of time at the observer. We assume a transient class emitting from $z\sim 0$, characterised by the unrealistic rate $1$ Hz and a fixed triangular light curve. The flux limit $F_{lim}$ is represented by the horizontal line. The transient light curves above the flux limit are shown in black solid lines. The light gray region highlights the 10~s exposure window. 
The panels show the peak contribution, i.e. the events whose luminosity peak falls in the observational window (green stars) and the tail contribution, i.e. the events detected only before or after the luminosity peak (fuchsia lines and area). The only difference between upper and lower panels is the duration of the transient light curve above the flux limit, respectively $1$~s and $2$~s.
Doubling the light curve duration results in doubling the number of tails, while leaving unchanged the number of peaks.
}
            \label{fig:peaks_tails}
  		\end{minipage}
\end{figure}

%%%%%%%%%%%%%%%%%%%%%%%%%%%%%%%%%%%%%%%%%%%%%%%%%%%
\texttt{saprEMo} can be applied to any type of EM emission, from transients to continuous sources
emitting in any EM spectral range. 
In this work,
we focus on X-ray transients associated with mergers of neutron star binaries.
The expected number of BNS mergers $N_{BNS}$ in the volume enclosed within redshift $z_{max}$, in a time $T$ at the observer, is:
\begin{eqnarray}
N_{BNS}=T\int_{0}^{z_{max}}\frac{R_V(z)}{1+z}\frac{dV_C}{dz}\,dz
\label{eq:nBNS}
\end{eqnarray}
where $R_V(z)$ is the rate of BNS mergers per unit comoving volume, per unit source time. In our case $z_{max}$ is the maximum distance at which the emission following the model of interest {\it EMe}, can be detected.
$z_{max}$ is calculated considering both the spectral shift due to the source redshift compared to the instrument energy band $E^I\sim\left[E^I_{min},E^I_{max}\right]$ and the maximum luminosity distance, set by the peak luminosity $L_p(E)$ of the {\it EMe} model
and the sensitivity $F_{lim}$ of the survey.
We only expect a fraction of $N_{BNS}$ to be observed
by a specific instrument, depending on the emission properties and the characteristics of the survey.
The number of BNS mergers, detectable by the survey {\it S} such that the peak of the considered emission {\it EMe} falls within the observing time, is given by:

\begin{eqnarray}
\begin{split}
N_{p}=
 \varepsilon~ \frac{FoV}{4\pi} T \int_{0}^{z_{max}} \frac{\displaystyle R_V(z)}{1+z}\frac{dVc}{dz}\,dz
\label{eq:peaks}
\end{split}
\end{eqnarray}
The total observing time $T$ can be expressed in terms of the survey {\it S} as $T = \left<T_{obs}\right>~n_{obs}$, where $n_{obs}$ is the number of the observations and 
$\left<T_{obs}\right>$ is the average exposure time for observation.
In eq. (\ref{eq:peaks}), the field of view $FoV$ of the instrument reduces the number $N_{BNS}$ of signals present in the volume enclosed within $z_{max}$ by $FoV/4\pi$.
In the specific case of BNS mergers, the efficiency factor $\varepsilon$ typically includes the 
occurrence rate of a specific merger remnant ($\varepsilon_{sr}$), which are expected to generate the emission {\it EMe}, and source geometry/observational  restrictions such as collimation ($\varepsilon_c = 1-\cos(\theta)$, where $\theta$ is the beaming angle), so that $\varepsilon\sim \varepsilon_{sr}\cdot \varepsilon_c$.
We designate as {\it peaks} the signals included in $N_p$ (see figure \ref{fig:1}).
This contribution only depends on the emission model by the intensity of the light curve peaks in the energy bands of the survey. This dependency is enclosed in $z_{max}$.

There is also a contribution, which we call {\it tails} (figure \ref{fig:1}), from the mergers whose emission is detected only before (first block of eq. (\ref{eq:tails})) or after (second block of eq. (\ref{eq:tails})) the luminosity peaks (i.e., $L_p$ is outside the observation period). The longer the light curve is above $F_{lim}$, the higher the probability of it being observed (see \citealt{carbone2016calculating} for a detailed discussion of signal duration in the context of transient detectability and classification).
To estimate this contribution, we need to account for the evolution in time of the emission luminosity $L(t')$, which affects the horizon of the survey:

\begin{eqnarray}
\begin{split}
N_t & =\displaystyle \varepsilon~ n_{obs}\frac{FoV}{4\pi}\left[\int_{-\infty}^{t'_p}\int_{0}^{z_t(L(t'))} \frac{\displaystyle R_V(z)}{1+z}\frac{dVc}{dz}\,dz\,dt +\right.\\ & \quad \left.\int_{t'_p}^{+\infty}\int_{0}^{z_t(L(t'))} \frac{\displaystyle R_V(z)}{1+z}\frac{dVc}{dz}\,dz\,dt \right] \\ & =\displaystyle \varepsilon~ n_{obs}\frac{FoV}{4\pi}\int_{-\infty}^{+\infty}\int_{0}^{z_t(L(t'))} \frac{\displaystyle R_V(z)}{1+z}\frac{dVc}{dz}\,dz\,dt
\label{eq:tails}
\end{split}
\end{eqnarray}
where $t$ and $t' = t/(1+z)$ are the time respectively in observer and source frames and $t'_p$ is the time corresponding to the peak luminosity. $z_t\left(L(t')\right)$ represents the horizon of the survey, given the specific intrinsic luminosity of the source $L(t')$. 
The integration time of eq. (\ref{eq:tails}) is practically limited by the duration of the emission above the flux limit.
At the moment, \texttt{saprEMo} does not correctly account for the possibility of 
detecting multiple times
the same event. Multiple detections of the same source might occur if the survey contains repeated observations of the same sky locations and the time interval between the different exposures is 
shorter than
the considered emission {\it EMe}. 
While $N_p$ would be unaffected, in these cases $N_t$ would overestimate the expected number of events by these additional detections. Under these specific conditions, $N_t$ should then be considered as an upper limit.
We refer the readers to \citealt{carbone2016calculating} for discussions of transient detectability in the context of multiple images of the same field.

%%%%%%%%%%%%%%%%%%%%%%%%%%%%%%
%%%%% 
The ratio between the duration of the observable emission and the typical exposure time determines the relative importance of peaks and tails.
The trade off between these two contributions, as well as their different origin, can be understood with figures \ref{fig:1} and \ref{fig:peaks_tails}.  
For illustrative purposes, we consider  only local events with unphysical rates and a generic triangular light curve. 
Figure \ref{fig:1} shows how tails events $b$ and $c$ can be observed during one exposure of duration $\left<T_{obs}\right>$.
Figure \ref{fig:peaks_tails} shows the impact of the transient observable duration on tails. Upper and lower panels represent exactly the same scenarios (10 seconds of exposure of transients at $z = 0$ characterised by a 1 Hz rate) except for the transient duration, which is doubled in the lower panel. The number of stars, which represent the peak contribution, is the same in both upper and lower panels, demonstrating that peaks are unaffected by the change of the transient observable duration. 
However the tail contribution, given by the number of pink triangles, doubles in the lower panel compared to the upper one.
On the contrary, extending the exposure to 20 seconds would double the number of peaks, while leaving unchanged the number of tails. 
Given a fixed event rate, $N_p$ depends on the observing time, while $N_t$ depends on the duration of the events.
We return to this topic in section \ref{S4:Discussion}, when we discuss the results of this study.

While eq. (\ref{eq:peaks}) and (\ref{eq:tails}) explain the general concept behind the tool, they do not explicitly account for the energy (or wavelength $E=\hbar c/\lambda$) dependence of light curves, instrument sensitivities and absorption.
These effects are particularly important when exploring the Universe at high redshift. 
We explain how these effects are implemented in \texttt{saprEMo} analysis in appendix \ref{App:A}.\\

%%%%%%%%%%%%%%%%%%%%%%%%%%%%%%

%%%%%%%%%%%%%%%%%%%%%%%%%%%%%%%%%%%%%%%%%%%%%%%%%%%
\subsection{Inputs and Outputs}
%%%%%%%%%%%%%%%%%%%%%%%%%%%%%%%%%%%%%%%%%%%%%%%%%%%

We now present inputs and outputs of
\texttt{saprEMo}.\\

\noindent {\it Inputs:}
\begin{enumerate}[i]
\item \textbf{light curves} (LC), in terms of luminosity rest frame, of the EM emission {\it EMe} in different energy bins 
(if predicted by the model, also including energies outside the instrument sensitivity band, as they 
might be redshifted into it, after accounting for cosmology corrections); 
{\es{(spin-down luminosity of the stable newly born neutron star - fiducial light curve proposed by \citealt{Siegel2016a})}}
\item \textbf{astrophysical rate} in the source frame $R_V\left(z\right)$;
{\es{( rate of BNS mergers, as derived by \citealt{Domink2013})}}
\item \textbf{efficiency} $\boldsymbol{\varepsilon}$ of the EM model {\it EMe} ($\boldsymbol{\varepsilon}$ should account for source geometry/selection effects, such as collimation, as well as the frequency with which this type of astrophysical source generates the electromagnetic emission {\it EMe}); 
{\es{(assuming this corresponds to the fraction of SGRBs followed by extended X-ray emission $\varepsilon\sim\%50$ \citealt{Rowlinson2013})}}
\item  \textbf{main instrument and survey properties}:
\begin{itemize}
\item for each spectral band $g$ of the survey {\it S}, minimum and maximum energy included $[E_i,E_s]_g$;
\item corresponding flux limits $[F_{lim}]_g$;
\item average exposure time $\left<T_{obs}\right>$;
\item field of view $FoV$ \footnote{Or equivalently covered sky-area $f_{sky}\sim n_{obs}\frac{FOV}{4\pi}$.};
\item number of observations $n_{obs}$\footnotemark[1].
\end{itemize}
{\es{({XMM-Newton Serendipitous Source Catalog. Averaged sensitivity {\ToBeChecked{$\sim 10^{-15}$\,\flux}}, average observation time {\ToBeChecked{$\sim 10^{4}$\,s}}, instrument band $0.2-12$\,kHz, field of view $\sim 0.2$\,$deg^2$})}}
\end{enumerate}

\noindent {\it Outputs:}
\begin{itemize}
\item $\boldsymbol{\mathrm{N_p}}$ and $\boldsymbol{\mathrm{N_t}}$: numbers of peaks and tails which are expected in the survey {\it S}. The numbers of signals returned by \texttt{saprEMo} should be interpreted as the expectation value of a Poisson process. 
Therefore Poisson statistical errors should be considered in addition to the systematics due to rate and emission model uncertainties;
\item $\boldsymbol{\mathrm{dN_p}}${\bf/dz} and $\boldsymbol{\mathrm{dN_t}}${\bf/dz}: distributions of tail and peak numbers as a function of redshift;
\item $\boldsymbol{\mathrm{dN_p}}${\bf/dlog(D)} and $\boldsymbol{\mathrm{dN_t}}${\bf/dlog(D)}: expected distribution of signal observed durations, obtained by convolving the approximate distribution of the survey exposure times $P_{obs}$ with the light curve span $LCS$ observable at each step in redshift.
For each redshift, the $LCS$ represents the total time of the light curve which is above the flux limits at the observer frame (for more details see appendix \ref{APP_A3}).
To estimate the distribution of the signal durations, 
we analytically approximate the exposure time distribution $P_{obs}$ from the average exposure time $\left<T_{obs}\right>$ (and standard deviation, when available) with a {\it Maxwell-Boltzmann} or {\it Log-normal} function, according to the user input. 
For each data point saved from the cosmic integrations, 
we simulate $N_{trials}$ (for both peaks and tails)
observation durations $T_{obs}$ and define
for each of them the starting time $t_s$.
The starting time is uniformly drawn from a time interval including both the exposure time of the specific trial $T_{obs}$ and the observable emission at observer $(t'_f - t'_i)(1+z)$ (where $t'_f$ and $t'_i$ are respectively the last and first LC time at source satisfying our detection criteria at redshift $z$).
If $t_s$ is drawn in the interval $I_p =\left[t_p - T_{obs}, t_p\right]$, where $t_p$ is the time at observer correspondent to the luminosity peak, it contributes to the peak distribution, otherwise it adds up to the tail distribution. For each simulated observation the total duration is then calculated summing only the contribution of light curve intervals whose flux is above the limit;
\item $\boldsymbol{\mathrm{dN_p}}${\bf/dlog(F)} and $\boldsymbol{\mathrm{dN_t}}${\bf/dlog(F)}: distributions of peak and tail detection numbers as a function of maximum flux. At each step in redshift, necessary to compute the integral (\ref{eq:peaks}), \texttt{saprEMo} also calculates the associated flux. 
The fluxes are obtained by summing the contribution of each energy and rescaling with the associated luminosity distance. From the same observations simulated for estimating the duration distributions, we obtain the expected distribution of maximum fluxes.
\end{itemize}
Distributions in redshift are useful to estimate the horizon of the survey to the emission {\it EMe} and for astrophysical interpretation. 
They provide a prior on the redshift distribution when a counterpart allowing $z$ measurement is missing, or constrain cosmic rate evolution of BNS mergers when multi-wavelength observations yield the source distance.
Distributions of fluxes and durations are robust observables, which 
enable comparisons with real data
\footnote{
The reported flux distribution is calculated from the maximum theoretical fluxes of detected events in our simulation (see paragraph on $dN_{p/t}/d\log(D)$-output). 
Quantitative comparisons with actual data would require a more detailed analysis, including 
the use of the instrument response, a realistic model for noise, the model used to convert photon counts into a   light curve, etc.~(see for example \citealt{carbone2016calculating}, who modeled some of these aspects).}.

\section{Application to soft X-ray emission from long-lived binary neutron star merger remnants}
\label{S3:Xray_sky}
%%%%%%%%%%%%%%%%%%%%%%%%%%%%%%%%%%%%%%%%%%%%%%%%%%%
In the following, we consider a specific application of \texttt{saprEMo} to the case of spindown-powered X-ray transients from long-lived NS remnants of BNS mergers.

Depending on the involved masses and the NS equation of state (EOS), a BNS merger can either produce a short-lived remnant, collapsing to a black hole (BH) within a fraction of a second, or a long-lived massive NS. The latter can survive for much longer spindown timescales (up to minutes, hours or more) prior to collapse or even be stable forever \citep{lasky2014nuclear, Lu2015}. 
After the discovery of NSs with a mass of $\sim2~M_\odot$ \citep{Demorest2010, Antoniadis2013}, different authors converged to the idea that the fraction of BNS mergers leading to a long-lived NS remnants should range from a few percent up to more than half (e.g., \citealt{Piro2017}). 
Information extracted from the multimessenger observations of the BNS merger event GW170817 did not change this view, although more stringent constraints on the NS EOS were obtained from the GW signal \citep{LVC-EOS}, from various indications excluding the prompt collapse to a BH, and from the kilonova brightness and the relatively high mass of the merger ejecta (e.g., \citealt{margalit2017constraining,Bauswein2017,radice2018gw170817,Rezzolla2018}). 
An additional supporting element in favour of long-lived remnants is given by the observation of long-lasting ($\sim$ minutes to hours) X-ray transients following a significant fraction of SGRBs (e.g., \citealt{Rowlinson2013,Gompertz2014,Lu2015}). Given the short accretion timescale of a remnant disk onto the central BH ($\lesssim1$~s), such long-lasting emission represents a challenge for the canonical BH-disk scenario of SGRBs while it can be easily explained by alternative scenarios involving a long-lived NS central engine, e.g., the magnetar \citep{Zhang2001,Metzger2008} and the time-reversal \citep{Ciolfi2015,Ciolfi2018} scenarios. According to this view, the fraction of SGRBs accompanied by long-lasting X-ray transients might reflect the fraction of BNS mergers producing a long-lived NS.

If the merger remnant is a long-lived NS, its spindown-powered electromagnetic emission represents an additional energy reservoir that can potentially result in a detectable transient. Recent studies taking into account the reprocessing of this radiation across the baryon-polluted environment surrounding the merger site have shown that the resulting signal should peak at wavelengths between optical and soft X-rays, with luminosities in the range $10^{43}-10^{48}$~erg/s and time scales of minutes to days (e.g., \citealt{Yu2013,MetzgerPiro2014,siegel2016electromagnetic,siegel2016electromagneticII}). 
Besides representing an explanation for the long-lasting X-ray transients accompanying SGRBs, this spindown-powered emission is a promising counterpart to BNS mergers (e.g., \citealt{Stratta2017} and refs.~therein), having the advantage of being both very luminous and nearly isotropic.

For our first direct application of \texttt{saprEMo}, 
we consider the spindown-powered transient model by \citealt{siegel2016electromagnetic,siegel2016electromagneticII} (hereafter SC16), described in the next Section \ref{S3_1:Xray_model}, in which the emission is expected to peak in the soft X-ray band. 
This model cannot be excluded or constrained by GW170817. The first X-ray observations in the $2-10$~keV band were performed by MAXY \citep{sugita2018maxi} 4.6 hours after the merger with a sensitivity of $8.6\times10^{-9}~\mathrm{erg~s^{-1}~cm^{-2}}$,
well above the flux that the model predicts at that time after the merger.

In Section \ref{sec:3.2}, we briefly describe the model of the BNS merger rate adopted for this first study.
Then, in Section \ref{S3_3:Xray_results} we present our results referring to three different X-ray satellites: {\it XMM-Newton}, {\it Chandra}, and the proposed {\it THESEUS}. We discuss these results in section \ref{S4:Discussion}. 

%%%%%%%%%%%%%%%%%%%%%%%%%%%%%%%%%%%%%%%%%%%%%%%%%%%
\subsection{Reference emission model}
\label{S3_1:Xray_model}
%%%%%%%%%%%%%%%%%%%%%%%%%%%%%%%%%%%%%%%%%%%%%%%%%%%

The model proposed by Siegel \& Ciolfi (SC16) describes the evolution of the environment surrounding a long-lived NS formed as the result of a BNS merger. The spindown radiation from the NS injects energy into the system and interacts with the optically thick baryon-loaded wind ejected isotropically in the early post-merger phase, rapidly forming a baryon-free high-pressure cavity or ``nebula'' (with properties analogous to a pulsar wind nebula) surrounded by a spherical shell of ``ejecta'' heated and accelerated by the incoming radiation. 
As long as the ejecta remain optically thick, the non-thermal radiation from the nebula is reprocessed and thermalised before eventually escaping. As soon as the ejecta become optically thin, a signal rebrightening is expected, accompanied by a transition from dominantly thermal to non-thermal spectrum. The model can also take into account the collapse of the NS to a BH at any time during the spindown phase.\footnote{
We refer to \citealt{siegel2016electromagnetic,siegel2016electromagneticII} and \citealt{Ciolfi2016} for a detailed discussion of the model and its current limitations.}

Exploring a wide range of physical parameters, Siegel \& Ciolfi found that the escaping spindown-powered signal has a delayed onset of $\sim10-100$ s and in most cases peaks $\sim100-10^4$ s after merger. Furthermore, the emission typically falls inside the soft X-ray band (peaking at $\sim0.1-1$ keV) and the peak luminosity is in the range $10^{46}-10^{48}$ erg s$^{-1}$. 
\begin{figure}
\begin{centering}
\includegraphics[width=8.75cm] {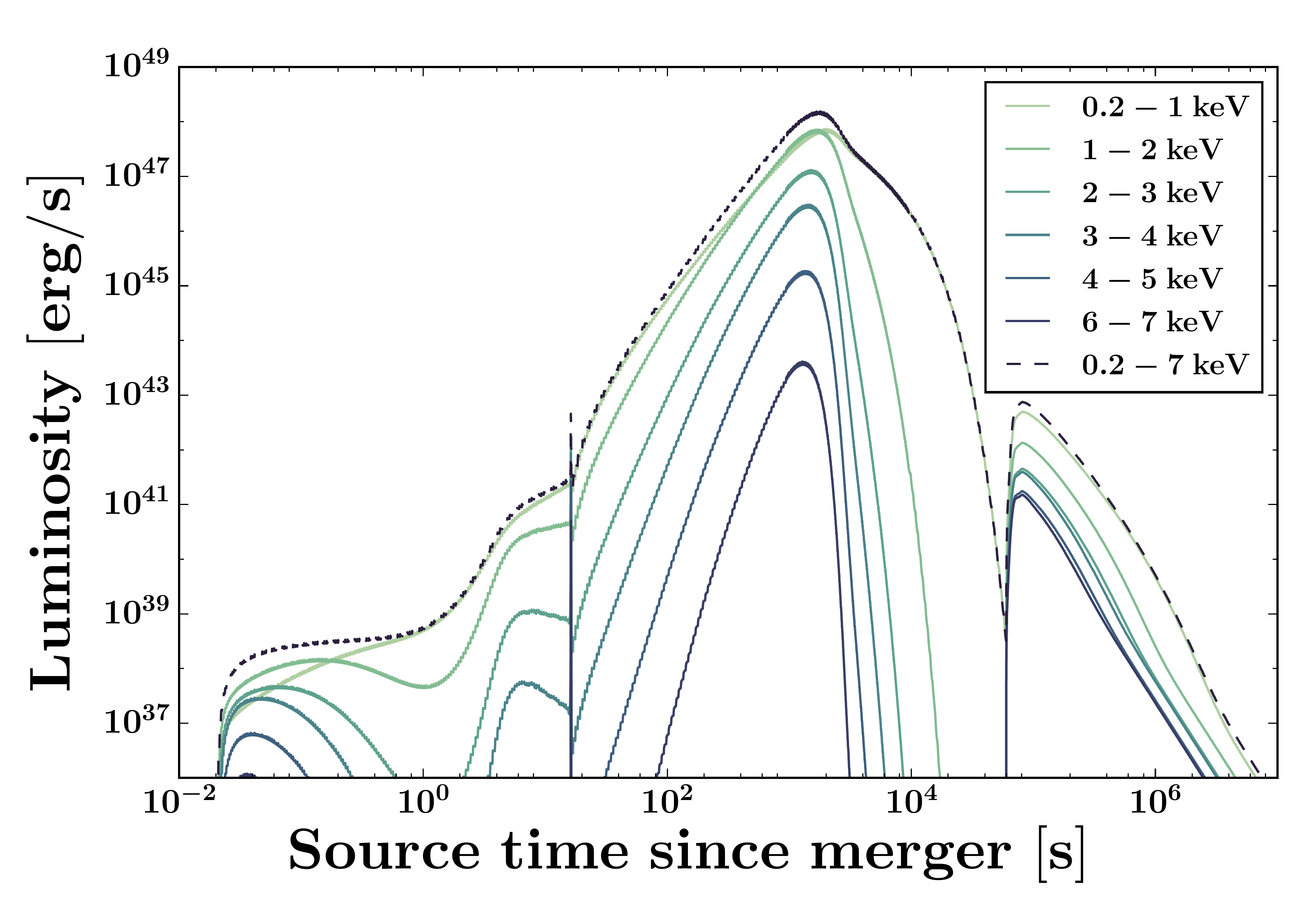}
\caption{Light curve of the spindown-powered emission from a long-lived BNS merger remnant according to the model proposed by \citealt{siegel2016electromagnetic,siegel2016electromagneticII} (corresponding to their ``fiducial'' case; see text).
The solid curves represent the contributions of different energy bands to the total light curve (dashed line).}
\label{fig:LCinE}
\end{centering}
\end{figure}
In this work, we consider only one representative case, corresponding to the ``fiducial case'' of SC16 (SC16f) (this model is rescaled for the
analysis in Section \ref{sec3.3.2}).
The light curve and spectral distribution of this particular model are shown in figure~\ref{fig:LCinE}. The main parameters of the model are as follows. The early baryon-loaded wind ejects mass isotropically at an initial rate of $5\times10^{-3}~M_\odot$~s$^{-1}$, decreasing in time with a timescale of 0.5~s. The dipolar magnetic field strength at the poles of the NS is $10^{15}$~G and the initial rotational energy of the NS is $5\times10^{52}$~erg ($\sim$~ms initial spin period).
Moreover, in this case the remnant evolves without collapsing to a BH.  
In figure~\ref{fig:LCinE} we can distinguish two important transitions. The first, around $\sim10$~s, marks the end of the early baryon wind phase and the beginning of the spindown phase. The second, at several times $10^{4}$~s, corresponds to the time when the ejecta become optically thin.  
While the emission described by the above model is essentially isotropic, allowing us to set $\varepsilon_c\sim 1$, only a fraction of BNS mergers $\varepsilon_{LLNS}$ is expected to generate a long-lived neutron star.
The value of this fraction mainly depends on the unknown 
NS EOS and distribution of component masses. Here, we assume for simplicity a one-to-one correspondence between the fraction $\varepsilon_{LLNS}$ and the fraction of SGRBs accompanied by a long-lasting X-ray transient (i.e. extended emission and/or X-ray plateau). Following the analysis presented in 
\citealt{Rowlinson2013}, we set $\varepsilon_{LLNS}$ to 50\%.

Once we assign $\varepsilon_{sr}=\varepsilon_{LLNS}$, the resulting total efficiency of the emission is $\varepsilon\sim \varepsilon_{sr}\cdot \varepsilon_{c}=50\%$.

\subsection{BNS merger rate model}
\label{sec:3.2}
\begin{figure}
\begin{centering}
\includegraphics[width=9.25cm]{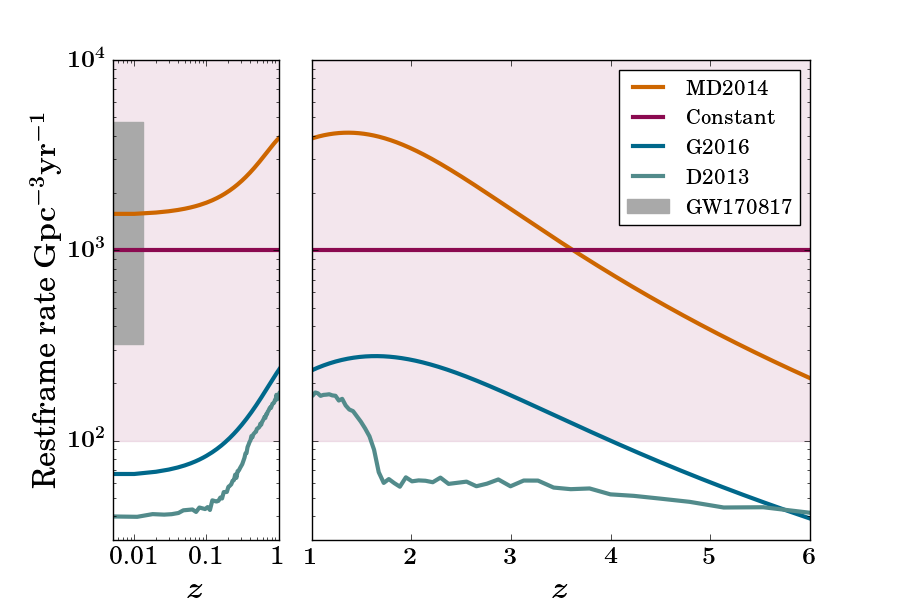}
\caption{BNS merger rate as a function of redshift for different models: D2013 \citealt{dominik2013double} in {green}, G2016 \citealt{ghirlanda2016short} in {blue}, MD2014 \citealt{madau2014cosmic} convolved with $P(t_{del})\propto t_{del}^{-1}$ in {orange} and our default constant model in {violet}: the solid line represents the reference rate of $R_V(z) = 1000 ~\mathrm{Gpc^{-3}yr^{-1}}$, while the light shadowed area includes the whole interval $R_V(z) =[100-10000]~\mathrm{Gpc^{-3}yr^{-1}}$.
The range in redshift is divided into $z\le 1$ and $z>1$ to enhance the visibility of the constraints set by the observation of GW170817 (gray area, 90\% probability as reported in \citealt{GW170817}), which only apply to the local Universe.
}
\label{fig:rate}
\end{centering}
\end{figure}

%%%%%%%%%%%%%%%%%%%%%%%%%%%%%%%%%%%%%%%%%%%%%%%%%%%
The dependence of the BNS merger rate on redshift is poorly observationally constrained.  Several models based on different assumptions have been proposed.
For the present work we consider 4 different rate models, a simplified (default) case and three further astrophysically-motivated scenarios:
\begin{description}
\item[{\bf DEFAULT}]: a constant BNS merger rate per unit comoving volume per unit source time in the range $R_V(z) = [100-10000]~\mathrm{Gpc^{-3}yr^{-1}}$, extending up to a maximum redshift of $z = 6$;
\item[{\bf D2013}]: the Monte Carlo population synthesis model of \citealt{dominik2013double} (their cosmological standard model, high-end metallicity scenario \citealt{syntheticuniverse}.);
\item[{\bf G2016}]: the analytic approximation based on SGRB observations described in eq. (12) of \citealt{ghirlanda2016short} (adopting the average value of the parameter reported for case {\it a} with an opening angle of $4.5\deg$);
\item[{\bf MD2014}]: the analytic prescription for the star formation history proposed by \citealt{madau2014cosmic} convolved with a probability distribution of delay times between formation and merger given by the power low $P(t_{del})\propto t_{del}^{-1}$, with a minimum delay time of $20\times 10^{6}~$yr, normalised to the local BNS merger rate of $1540~$Gpc$^{-3}$yr$^{-1}$, as estimated with GW170817 ($1540^{+3200}_{-1220}~\mathrm{Gpc^{-3}yr^{-1}}$ median and uncertainties at 90\% probability,  
\citealt{GW170817}).
\end{description} 
The different BNS merger rates are reported in figure \ref{fig:rate}. We note that D2013 and G2016, as proposed, are inconsistent with the local rate range obtained from GW170817 (gray region).  
However both inferred rate from a single gravitational-wave observation and population synthesis models rely on poorly constrained astrophysical model assumptions and are therefore highly uncertain.
We adopt these rate models for illustrative purposes to test the impact of different redshift-dependent merger rates.\\

To investigate the impact of other inputs, we adopt the DEFAULT simplified model of constant cosmological rate as a reference case.
We report distributions and results for $R_V(z) = 1000 ~\mathrm{Gpc^{-3}yr^{-1}}$.
Because the considered $R_V(z)$ is constant, the results for the upper (lower) bound of the whole range  $R_V(z) = [100-10000] ~\mathrm{Gpc^{-3}yr^{-1}}$, can be obtained by scaling up (down) the output quantities by one order of magnitude.
This wide range of the BNS merger rate includes the local rate interval inferred from the detection of GW170817 and is broadly consistent with estimates obtained using Galactic BNS observations and population synthesis models
(\citealt{abadie2010predictions,paul2017binary,chruslinska2017double,vignagomez2018}). 

We use the DEFAULT model in Section \ref{S3_3:Xray_results} and discuss the impact of applying different BNS merger rate models in Section \ref{S4:Discussion}.
%%%%%%%%%%%%%%%%%%%%%%%%%%%%%%%%%%%%%%%%%%%%%%%%%%%
\subsection{Results}
\label{S3_3:Xray_results}
%%%%%%%%%%%%%%%%%%%%%%%%%%%%%%%%%%%%%%%%%%%%%%%%%%%
We now proceed with using \texttt{saprEMo} and the X-ray transient model described in Section \ref{S3_1:Xray_model} to evaluate the expected number of detections of this type of signal in three present and future surveys by {\it XMM-Newton}, {\it Chandra} and {\it THESEUS}. 
To emphasise the impact of the survey properties, we initially keep fixed: (i) the light curve to the SC16 model described in Section \ref{S3_1:Xray_model} (fiducial, SD16f, or fiducial-rescaled, see Section \ref{sec3.3.2}); (ii) the assumed astrophysical rate to the DEFAULT case, described in Section \ref{sec:3.2}; and (iii) the efficiency $\varepsilon\sim 50\%$.

In particular, we consider:
\begin{itemize}
\item two present surveys, collected during the decade of operation of {\it XMM-Newton}, to predict the expected number of detectable signals in these archived data;
\item the {\it Chandra} Deep Field - South (CDF-S)
data set to verify whether the transient class discovered by \citealt{bauer2017new} is statistically consistent with the SC16 model;
\item 10~ks of {\it THESEUS} observations, to explore 
the sensitivity of this mission concept
to transients associated with BNS mergers, such as SC16f.
\end{itemize}
The main properties of the surveys are summarised in appendix \ref{App:Survey_Prop}.
%%%%%%%%%%%%%%%%%%%%%%%%%%%%%%%%%%%%%%%%%%%%%%%%%%%
%%%%%%%%%%%%%%%%%%%%%%%%%%%%%%%%%%%%%%%%%%%%%%%%%%%
\subsubsection{XMM-Newton}
\label{sec3.3.1}
\begin{figure*}
    \centering
    \subfloat[][]
    {
        \includegraphics[scale=.365]{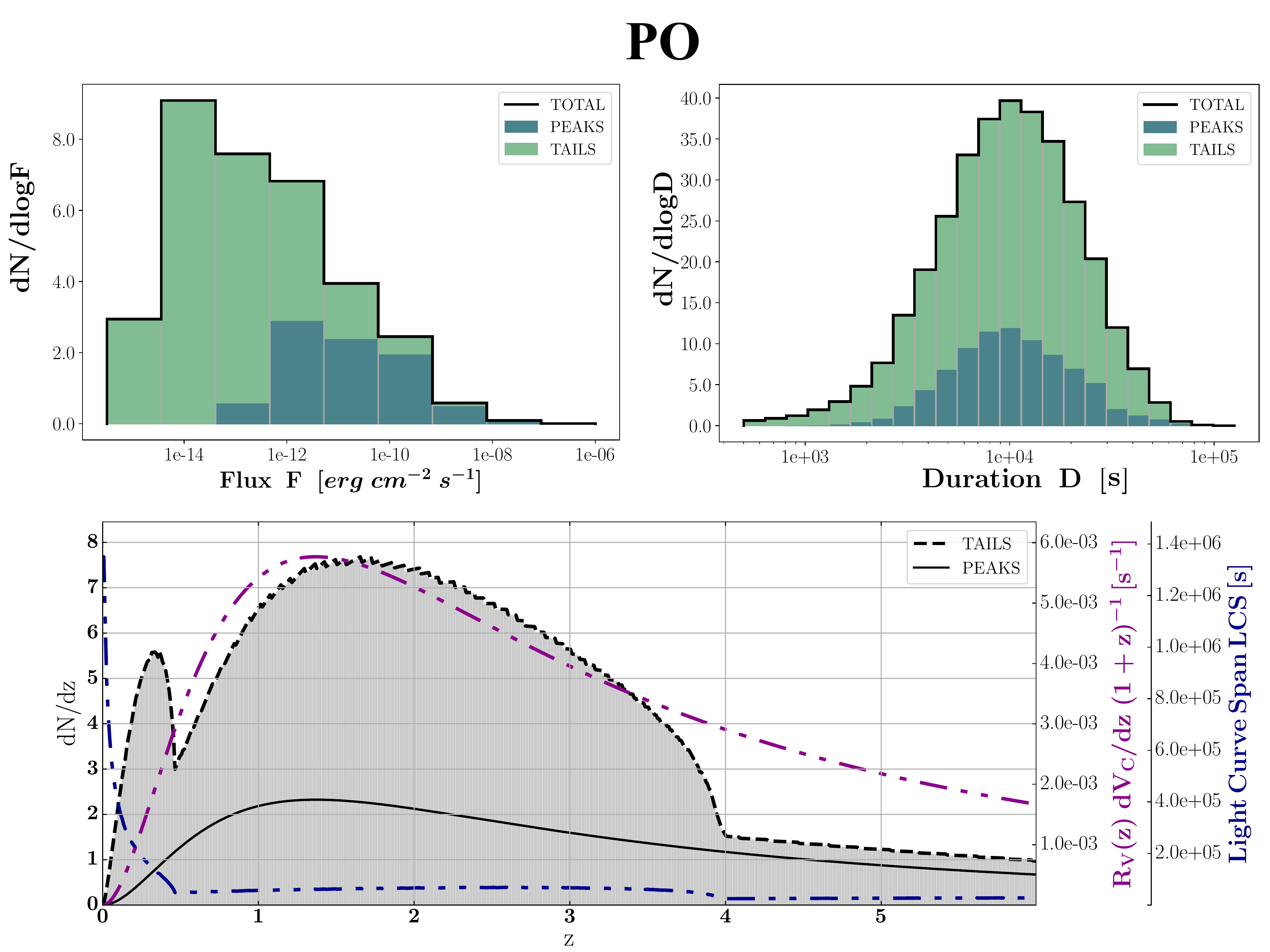}
    }
    \qquad
    \subfloat[][]
    {
        \includegraphics[scale=.365]{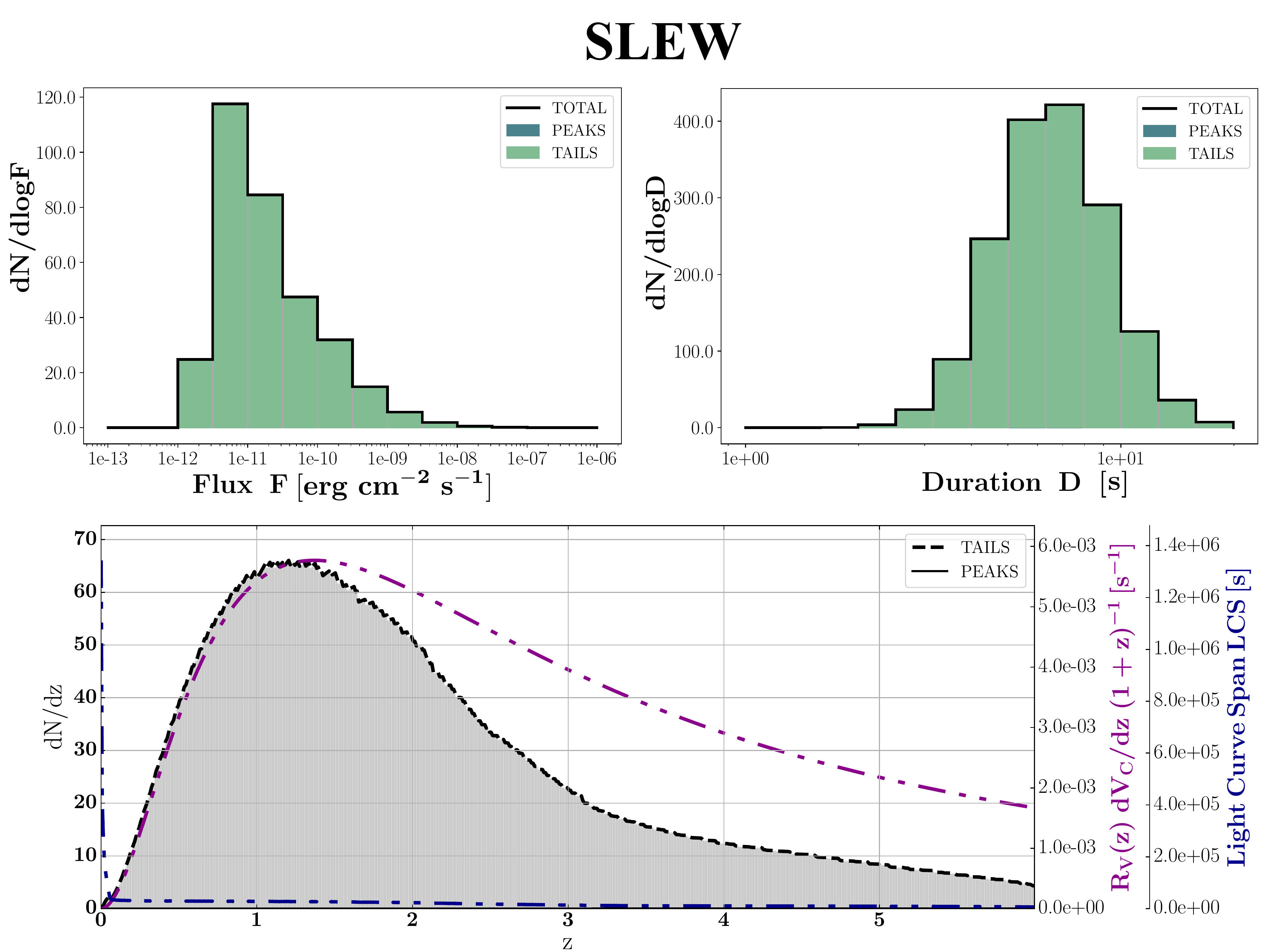}
    }
    \caption
    {Comparison between expected distributions of detectable signals in PO and SLEW surveys. The upper-left panel shows the total expectations on the observed maximum flux distribution obtained by adding peak (dark green) and tail (green) contributions. The upper-right panel shows the expected distributions of peak and tail durations. The redshift distributions in the bottom panel represent the differential contribution of peaks and tails throughout the scanned comoving volume of Universe. The violet and blue curves have been added to explain the trend of the tail distribution and represent respectively the time-shifted rate of events per unit redshift (scaled argument of the $z$ integral \ref{eq:tails_rev}) and the light curve span $LCS(z)$ (the duration of the light curve above the flux limits at a fixed $z$).
        (a) Expected distributions of observed maximum flux (upper-left), duration of signals above the flux limits (upper-right) and redshift (bottom) of the observations in {\it XMM-Newton} PO survey.
        (b) Expected distributions of observed maximum flux (upper-left), duration of signals above the flux limits (upper-right) and redshift (bottom) of the observations in {\it XMM-Newton} SLEW survey. Because of the low expected number of detections for peaks compared to tails (see tab. \ref{tab:XMM_res}), the black solid line in bottom graph is not visible.
    }
    \label{fig:XMMres}
\end{figure*}

We apply \texttt{saprEMo} to two different
collections of data obtained by 
{\it XMM-Newton}; we call them SLEW and PO (Pointed Observations), their characteristics are presented in the following.
The number of signals predicted by \texttt{saprEMo} are reported 
in table \ref{tab:XMM_res}
\footnote{For both the surveys, the statistical uncertainties due to the assumed Poisson distribution 
are almost negligible compared to the systematics due to 
uncertainty in the signal production efficiency $\varepsilon$ and the BNS merger rate.}.
In the case of {\it XMM-Newton} surveys, the sky locations of observations have been used to estimate the impact of the absorption due to the Milky Way (see appendix \ref{APP_A2} for more details on our adopted absorption model).
\begin{description}
\begin{table}
\centering
\begin{tabular}{|c|c|c|c|c|c|}
&\multicolumn{2}{|c|}{{\bf {\it XMM-Newton}}} &{\bf {\it Chandra}} &\multicolumn{2}{|c|}{{\bf {\it THESEUS}}}\\
\hline
 & PO & SLEW &CDF-S & Case a& Case b\\
\hline
$\mathbf{N_p}$ & 8 & 0& 0.14 & 5 (4) & 3 (2) \\
\hline
$\mathbf{N_t}$& 25 & 165 &\ding{56} & 5 (3) & 20 (11) \\
\hline
\hline
$\mathbf{FoV}~[\deg^2]$ & \multicolumn{2}{|c|}{$\sim 0.2$} & 0.08  & \multicolumn{2}{|c|}{{3300}} \\
\hline
$\mathbf{T}~[10^6\mathrm{s}]$ & $\sim 160$ & $\sim 1.06$ & $\sim6.73$ & \multicolumn{2}{|c|}{{$0.01$}}\\
\hline
\multicolumn{4}{c}{{}}\\
\\
\end{tabular}
\caption{
Average expected values for peaks ($N_p$) and tails ($N_t$) for different surveys. {\it XMM-Newton} PO and SLEW surveys$^a$, {\it Chandra} Deep Field - South (CDF-S) and 10 ks of {\it THESEUS} operation for a single exposure, {\it case a}, and 10 distinct exposures {\it case b} considering $N_H = 5\times 10^{-22}~\mathrm{cm^{-2}}$ ($N_H = 5\times 10^{-20}~\mathrm{cm^{-2}}$). $^a$ For the {\it XMM-Newton} surveys the total observing time was inferred from $T = n_{obs}\left<T_{obs}\right> = \left(4\pi~f_{sky}~FoV^{-1}\right)\left<T_{obs}\right>$, using the properties reported in tables \ref{tab:XMM_PO_par} and \ref{tab:XMM_SLEW_par}.
}
\label{tab:XMM_res}
\end{table}
\item[{\bf{The PO survey}}] is a collection of pointed observations made between 3/2/2000 and 15/12/2016. 
The data belong to the {\it XMM-Newton Serendipitous Source Catalog (3XMM DR7)}  (\citealt{3XMM-DR7ref,3XMM_Pref}). \\
PO exposures are longer (typically $10^3-10^4$~s) compared to the SLEW catalog (see following paragraph). 
This implies an extension to lower fluxes (down to $10^{-15}-10^{-16}~\mathrm{erg~s^{-1}~cm^{-2}}$), as figure \ref{fig:XMMres} (a) shows.
The same figure shows that such low fluxes are however reached only by tails.
This is because the luminosity of the model makes the flux higher than the survey flux limits up to $z=6$, which is the artificial cut of our BNS merger rate.
The distribution of source redshift is represented in the bottom graph of figure \ref{fig:XMMres} (a) and implies that, under the assumption of a constant cosmic BNS merger rate,
the median redshift of detectable signals is $z\sim 2$.
The double bump in the tail distribution of the PO survey is explained by the blue and purple curves which respectively represent the light curve span above the threshold at a fixed redshift $z$, $LCS(z)$, 
and 
the time-shifted rate of events per unit redshift, $\displaystyle R_V(z)\,dVc/dz$. 
Given our simplified BNS merger rate model,
the purple and black-solid lines scale like the 
redshift derivative of the comoving volume,
since in both cases only constants multiply the element $dV_c/dz$.
The blue curve has instead a very peculiar behaviour which depends on the specific emission light curve compared to the limit fluxes of the survey.
The distinct trends in the blue lines of figure \ref{fig:XMMres}, are due to features very peculiar to the adopted light curve (figure \ref{fig:LCinE}).
When the flux from the non-thermal second peak drops below the limit, the overall visible duration sharply decreases; this happens at $z\sim 0.5$ and $z\sim 0.05$ for PO and SLEW, respectively.
The second turn-over  at $z=4$ in the LCS, evident in PO tails (gray area and dashed-black line), is instead due to the discretisation of the light curves in energy bins and the relatively soft energy spectrum of these transients. 
In particular the lowest energy bin characterising the light curve (see figure \ref{fig:LCinE}) exits the band of the instrument \footnote{According to appendix \ref{App:A} notation: $z_{exit} =[E'_{max,h=0}/E^I_{min}-1]\sim [1~\mathrm{keV}/\mathrm{0.2}~\mathrm{keV}-1] = 4$, where $z_{exit}$ is the redshift at which the energy bin of lowest energy, denoted with $h = 0$, exits the instrument band, $E'_{max,h=0} = 1~keV$ is the highest energy included in the bin $h = 0$ and $E^I_{min} = 0.2~$keV is the minimum energy included in the instrument band.}.
\item[{\bf The SLEW survey}] is composed by data collected while changing the target in the sky, according to the {\it XMM-Newton} observation program (\citealt{SLEWref}).
The tested observations are collected in the 
{\it XMM-Newton Slew Survey Clean Source Catalog, v2.0}. \\
The SLEW survey is characterised by typical exposure time of only few seconds, and consequent flux limits $\gtrsim 10^{-13}~\mathrm{erg~s^{-1}~cm^{-2}}$.
Given the properties of the model SC16, this yields a predominance of tails over peaks (see first point of section \ref{S4:Discussion}). 
Our results show that SLEW observations, assuming correct identification (see section \ref{S4:Discussion}), could already reveal a population of BNS merger events.
The flux limits of the survey determines the distance of most of the X-ray sources at $z<3$.
\end{description}

Although the data have been collected by the same instrument, PO and SLEW considerably differ in terms of exposure time (and therefore sensitivity), sky coverage and energy responses (as shown by tables in appendix \ref{Appendix:XMM-Newton}). 
The SLEW survey is less sensitive, but it scans a much wider area of the sky compared to the PO survey, so that the total number of expected signals is actually considerably larger (see  tab. \ref{tab:XMM_res}). 
%%%%%%%%%%%%%%%%%%%%%%%%%%%%%%%%%%%%%%%%%%%%%%%%%%%  
\subsubsection{Chandra and the new, faint X-ray population}
\label{sec3.3.2}
%%%%%%%%%%%%%%%%%%%%%%%%%%%%%%%%%%%%%%%%%%%%%%%%%%%
\citet{bauer2017new} have recently claimed the discovery of a X-ray signal, belonging to a new, previously unobserved, transient class. The event was observed within the {\it Chandra} Deep Field - South (CDF-S), a deep survey of a 0.11 square degrees sky region composed of 102 observations collected in different periods during the last decade. 
Interestingly, the main properties of the event presented by \citet{bauer2017new} are broadly consistent with the SC16 emission model.
The maximum luminosity of $\sim 10^{47}~\mathrm{erg/s}$, the spectral peak around $\sim2$~keV (source frame), the rise time of $\sim 100~$s, and the overall duration of $\sim 10^{4}~\mathrm{s}$ are all in broad agreement with the model predictions. Here we do not attempt to provide convincing evidence for a potential match, but we want to show another interesting case for exploiting the capabilities of \texttt{saprEMo}. 
We assume a signal analogous to the SC16f adopted throughout this paper only rescaled to have a maximum luminosity of $\sim 10^{47}~\mathrm{erg/s}$ (referred to as ``rescaled SC16 model/signal'' in the following), consistent with 
the X-ray transient at $z\sim 2.23$ \citep{bauer2017new}.
To test the rate consistency between the detected X-ray transient and the rescaled SC16 model, we apply \texttt{saprEMo} to the CDF-S, adopting a Galactic neutral column density of $n_{H_{tot},MW} \approx n_{H,MW}\sim 8.8\times 10^{19}~\mathrm{cm^{-2}}$, as reported by \citet{bauer2017new}.
Given the proposed source redshifts, the shape and fluxes of the detected transient, 
we assume that the observed maximum flux corresponds to the luminosity peak of our model. 
We therefore 
evaluate only the expected number of peaks.
\texttt{saprEMo} predicts an expectation value of $\sim0.14$ signals in the CDF-S (see table \ref{tab:XMM_res}).
Given the adopted constant rate model, the probability of one rescaled SC16 signal being present at its luminosity peak in the $\sim 7~$Ms of the CDF-S is $\sim 12\%$ (with $\sim 87\%$ probability of $0$ signals). Considering the whole range of allowed BNS merger rates, this value ranges from $\sim 1.4\%$ (with $\sim 98.6\%$ probability of $0$ signals, correspondent to $R_V(z) = 100~\mathrm{Gpc^{-3}yr^{-1}}$) and $\sim 35\%$ (with $\sim 25\%$ probability of $0$ signals, correspondent to $R_V(z) = 10000~\mathrm{Gpc^{-3}yr^{-1}}$). 

Despite the broad consistency of the transient revealed by \citealt{bauer2017new} with the rescaled SC16 emission model,  
our analysis shows that a real association between the two is rather disfavoured, although not inconsistent given the uncertainties over rate and emission model.
Conversely, assuming that the detected transient is in fact the rescaled SC16 signal adopted in the above calculation, the constant BNS merger rate value 
is constrained to\footnote{To estimate the posterior on the constant rate value, we assume a flat prior in the range $[100,10000]~\mathrm{Gpc^{-3}yr^{-1}}$.} $6000^{+4000}_{-3700}~\mathrm{Gpc^{-3}~yr^{-1}}$ (median with 90\% credible interval, as in \citealt{GW170817}), higher than the median inferred from GW170817, though still consistent with the claimed interval
(cf.~Section \ref{sec:3.2}).
\subsubsection{Future observations with {\it THESEUS}}
\label{sec:3.3.3}
%%%%%%%%%%%%%%%%%%%%%%%%%%%%%%%%%%%%%%%%%%%%%%%
In the last few years, different wide-FoV X-ray missions have been proposed to monitor the X-ray sky, and specifically to follow up GRBs and GWs (\citealt{feng2016extp,barcons2012athena,yuan2015einstein,merloni2012erosita}). 
In particular, the mission concept 
 {\it THESEUS} has been recently selected by
ESA for assessment studies 
\citep{ESA_THESEUS} to explore the transient high-energy sky and contribute to multi-messenger astronomy \citep{Amati2017,Stratta2017,frontera2018observing}.
We apply \texttt{saprEMo} to test the sensitivity of the {\it THESEUS} mission to BNS mergers emitting in the X-ray according to the SC16f model.
On the {\it THESEUS} payload, the Soft X-ray Imager (SXI,  \citealt{o2018soft}) would be the instrument sensitive to such emission.
SXI flux sensitivities for sources in the Galactic plane ($N_H= 5 \times 10^{22}~\mathrm{cm^{-2}}$) and well outside it ($N_H = 5 \times 10^{20}~\mathrm{cm^{-2}}$) are taken from figure~4 of \citealt{Amati2017}.
With \texttt{saprEMo}, we predict detection numbers and properties for two cases
of gathering {\it THESEUS} observations, each having a total observing duration $T$ of 10~ks,
acquired with:
\begin{description}
\item[{\bf case a}] a single exposure of $\left<T_{obs}\right> = 10^4$~s;
\item[{\bf case b}] 10 exposures of non-overlapping sky regions, each lasting $\left<T_{obs}\right> = 10^3$~s.
\end{description}
\begin{table}
\centering
\begin{tabular}{|c|c|c|c|}
&\multicolumn{3}{|c|}{
{\bf {\it XMM-Newton} PO}} \\
\hline
{\bf Rate model }& MD2014 & D2013 & G2016 \\
\hline
$\mathbf{N_p}$&20 &1&2 \\
\hline
$\mathbf{N_t}$&65&2 &5\\
\hline
\hline
$\mathbf{FoV}~[\deg^2]$ & \multicolumn{3}{|c|}{{
$\sim$~0.25
}} \\
\hline
$\mathbf{n_{obs}\left<T_{obs}\right>}~[s]$ &\multicolumn{3}{|c|}{{$\sim 234\times 10^6~$}} \\
\hline
\\
\multicolumn{4}{c}{{}}\\
&\multicolumn{3}{|c|}{{\bf {\it THESEUS} Case {\it a}}} \\
\hline
{\bf Rate model }& MD2014 & D2013 & G2016 \\
\hline
$\mathbf{N_p}$& 17 (15) & 0.54 (0.46) & 1.2 (1.0)\\
\hline
$\mathbf{N_t}$& 16 (10) & 0.54 (0.34) & 1.0 (0.6)\\
\hline
\end{tabular}
\caption{
For {\it XMM-Newton} PO (top) and {\it THESEUS} case {\it a} surveys, comparison of expectation values assuming different BNS merger rate models, from left to right (i) analytic prescription proposed by \citealt{madau2014cosmic}, assuming a power-law distribution of delay times between formation and merger $P(t_{del})\propto t_{del}^{-1}$, (ii) cosmological rate derived by the population synthesis study \citealt{dominik2013double}, standard model at high-end metallicity scenario (D2013) and (iii) model based on SGRB statistics \citealt{ghirlanda2016short} with assumed opening angle of $4.5\deg$ (G2016).
}
\label{tab:RESULTS}
\end{table}

\begin{figure}
\centering
\includegraphics[width=9cm]{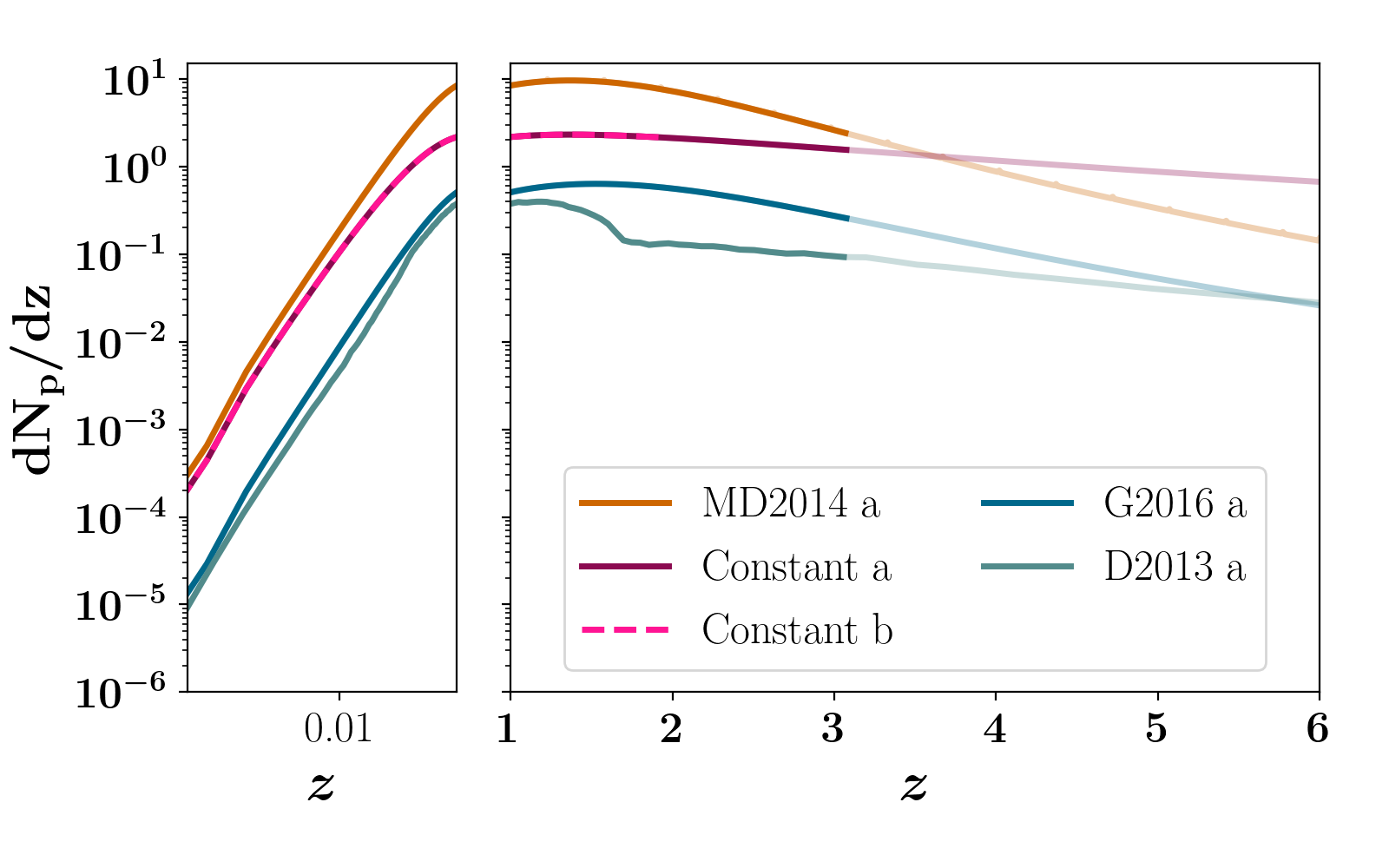}
\caption{Expected redshift distribution for peak signals in 10~ks of {\it THESEUS/SXI} observations (opaque lines), assuming $N_H = 5\times 10^{20}\mathrm{cm^{-2}}$. 
In the default configuration of constant BNS merger rates we test two different observing strategies: {\it case a} and {\it case b}, as described in Section \ref{sec:3.3.3}. They are shown by the solid-violet and dashed-fuchsia lines, which  completely overlap for $z \lesssim 2$ and only differ in the maximum achieved redshift, $z\sim 3$ for the former and $z\sim 2$ for the latter.
We also report the peak distribution as a function of redshift for other tested BNS merger models: MD2014 in orange, D2013 in green and G2016 in blue. For comparison with figure \ref{fig:rate}, we show the results in the two different regimes: $z>1$ in linear and $z\le 1$ in logarithmic scale. For the same reason, we also add in transparency
the distribution of peaks missed because of sensitivity constraints.
}
\label{fig:THESEUS1OBS_z} 
\end{figure}
The last 2 columns of table \ref{tab:XMM_res} show the expectation values for $N_p$ and $N_t$ 
for both case {\it a} and {\it b}.
{\it In less than 3 hours of total observing time T, we expect {\it THESEUS} to detect a number of SC16-like transients comparable to the ones inferred for years-long CDF-S {\it Chandra} and PO {\it XMM-Newton} surveys} (see also table \ref{tab:XMM_res}). 
The main reason why {\it THESEUS} is capable of reaching these numbers of detections in such a short observing time $T$ (about 4 orders of magnitude shorter than for CDF-S and PO/SLEW), is the 4 orders of
magnitude difference between its FoV and
the ones in {\it Chandra} and {\it XMM-Newton}. 
Albeit specific for the SC16f emission, our tabulated results demonstrate that the characteristics of the mission concept {\it THESEUS} suit the search for 
X-ray emission generated during BNS mergers. 
We predict that {\it THESEUS/SXI} with {\it case a} sensitivity would detect on the order of hundreds to thousands of BNS mergers in only a few years, assuming an emission with peak luminosity in the range $L_p\sim[10^{45}-10^{48}]~\mathrm{erg~cm^{-2}~s^{-1}}$ and a spectral distribution similar to SC16f.
Our analysis therefore shows that proposed large FoV instruments, such {\it THESEUS}, offer an incredible opportunity compared to present deep surveys, for the detection of rare but bright transients, such as those 
of the presented model. {\it THESEUS/SXI}-like campaigns are expected to detect events generated near the peak of the cosmic star formation and BNS merger rate, at $z\sim 1-3$. \\
The redshift distribution $dN_p/dz$ of case {\it a} and {\it b} in figure \ref{fig:THESEUS1OBS_z} show that, as expected, longer exposure times decrease the flux limits and therefore probe larger redshift.
Meanwhile, multiple shorter exposures at distinct sky locations, as in case {\it b}, increase the probability of detecting tails.  We find that {\it THESEUS/SXI} would make more detections, during a fixed total observing time, with an observing strategy that increased sky coverage at the cost of shorter exposures.

\section{Discussion}
\label{S4:Discussion}
%%%%%%%%%%%%%%%%%%%%%%%%%%%%%%%%%%%%%%%%%%%%%%%%%%%
With \texttt{saprEMo}, we tested the sensitivity of different astronomical surveys to the emission model SC16f.\\
\\
{\bf{
Given a fixed emission model, the tail contribution becomes more important for surveys with shorter exposures $\boldsymbol{\left<T_{obs}\right>}$.
}}
The analysis has confirmed that when the exposure $\left<T_{obs}\right>$ is considerably shorter than the duration of the emission model, as is the case of {\it XMM-Newton} SLEW observations, detections of pre-peak rises or post-peak decays will be more common than observations of peak flux.
Therefore the relation between $\left<T_{obs}\right>$ and the typical duration of the EM model sets the relative number of peaks and tails. 
Given a fixed amount of total observing time, the number of expected tails increases with the number of pointings of different sky positions. 
This is shown, for example, by the two cases tested with {\it THESEUS} (see tables \ref{tab:XMM_res} and \ref{tab:RESULTS}). \\
\\
{\bf{Inferences on BNS merger models}}
%\begin{changemargin}{0.5cm}{0.0cm} 
\begin{itemize}
\item{\bf{A few detections can already constrain the lower limit of cosmic BNS merger rate.}}
For example, assuming the emission model proposed by \citealt{siegel2016electromagneticII}, the probability of detecting more than 3 peaks in {\it XMM-Newton} PO
assuming a constant merger rate of $100~\mathrm{Gp^{-3}yr^{-1}}$
is $<1\%$; thus, a few detections could set a lower limit on the BNS merger rate.
The proposed emission model also offers the unique possibility of exploring mergers in the high-redshift Universe:
with peak luminosity as high as $\sim 10^{48}~ \mathrm{erg/s}$, {\it XMM-Newton} could detect 
signals generated at redshifts as high as $z\sim 15$ with PO sensitivities. \\
Constraints on BNS merger rate can be also obtained by assuming the association of the {\it Chandra} X-ray transient with the SC16 emission model. 
In this case our analysis can put a lower limit on the BNS merger rate of $\sim 2300~\mathrm{Gpc^{-3}yr^{-1}}$ at 90\% confidence interval, assuming a constant rate up to $z\sim 6$.
The peak luminosity of the rescaled SC16 emission is indeed bright enough to actually be detectable by {\it Chandra} up to $z\sim 6$, even after accounting for the Milky Way absorption.
\item{\bf{Larger PO-like or {\it THESEUS} surveys will probe $\mathbf{R_V(z)}$ and likely constrain also EM emission models.}}
In the case of bright emission, as expected by SC16, 
\texttt{saprEMo} predicts that the PO campaign can detect all BNS merger peaks in the Universe, localised within {\it XMM-Newton}'s field of view.
However, table \ref{tab:XMM_res} shows that to probe a statistically significant population of BNS mergers, PO-like sensitivities must be obtained with longer time window $T$ and/or larger FoV.
Large field of view instrument, such as {\it THESEUS/SXI}, can also detect many of these events.
Although limited to smaller redshift, cosmological distances could still be achieved if we assume bright emission models. This is shown for the case of SC16 in figure \ref{fig:THESEUS1OBS_z}.
These campaigns should therefore yield the BNS merger rate distribution as
a function of redshift, providing that multi-wavelength observations will allow redshift associations. These observations would then enable us to constrain the proposed BNS merger rate scenarios, which indeed predict different z distributions (see figure \ref{fig:THESEUS1OBS_z}).
\item{\bf{Redshift measurement play a fundamental role for breaking the degeneracy between emission parameters and rate models.}}
We apply \texttt{saprEMo} to the PO and {\it THESEUS} cases, {\it case a}, for the three scenarios
introduced in section \ref{sec:3.2}: D2013,  G2016 and MD2014 (see figure \ref{fig:rate}).
The absolute expectation values reported in table \ref{tab:RESULTS} and figure \ref{fig:THESEUS1OBS_z} , reflect the trend of the rate models reported in figure \ref{fig:rate}.
The overall results show that, without perfect knowledge of light curve and spectrum of the emission, measurements of source distances  are necessary to constrain the redshift dependence of the BNS merger rate.
\end{itemize}
{\bf{Considerations on \texttt{saprEMo}'s results.}} 
In this paragraph we highlight some general considerations, to realistically interpreting \texttt{saprEMo} results.
The main output of the analysis consists in the number of peaks and tails expected for a specific emission model in a selected survey of data.
However, depending on the purpose of the analysis, 
other information, such as more accurate requirements for detectability and classification, should be taken into account. In the following we give some examples.
\begin{description}
\item {\it Challenges for detectability:} 
despite this does not concern the results presented in the previous session, short transients in long exposure observations can be lost in the integrated background flux. 
To overcome this issue, targeted analyses might be required (see for e.g. the work of EXTraS group \citealt{extrasWEB_transient} \citealt{de2016science}
for detection of $\sim 10^2~$s-lasting transients in PO). 
\item {\it Challenges for classifications: }
because of their definition, we generally expect durations and fluxes of tails to extend to lower values compared to the peak ones (as shown in figure~\ref{fig:XMMres} (a)). 
This generally worsen the performances of signal classifications and identification among more common phenomena.
In particular for our analyses, given the shape of SC16f's light curve, tails should mostly appear as simple decaying signals, which can be challenging to distinguish from other X-ray transients (e.g., tails of tidal disruption events \citealt{Lodato2011} or supernovae \citealt{dwarkadas2011published}).\\
The correct classification of X-ray events can also be challenged by short exposure times.
This is for example the case of the SLEW survey, where observations typically last only few seconds.
Indeed some emission models, as SC16, predict a long-scale time evolution of the emission properties which would likely result in detections of dissimilar signals, challenging their association to a common origin.
Campaigns characterised by typically longer observations, such as the {\it XMM-Newton} PO and {\it Chandra} CDF-S, are less affected by classification problems.
The extension of the typical exposure time to thousands of seconds and improved spectral resolution, allow for the acquisition of more informative data,
simplifying transient identification.
\end{description}
\section{Summary and Outlook}
\label{S5:Conclusions}
%%%%%%%%%%%%%%%%%%%%%%%%%%%%%%%%%%%%%%%%%%%%%%%%%%%
In this study we showed some applications of our tool \texttt{saprEMo}; we applied it on few present and possible future surveys, assuming a specific emission model and cosmological BNS merger rate $R_V(z)$.
In terms of multi-messenger astronomy, our results show that the luminosities predicted by the SC16 emission model can be detected up to cosmological distances which extend much further than the horizon of present (\citealt{aasi2016prospects,2018LRR....21....3A}) and future gravitational wave detectors (\citealt{sathyaprakash2012scientific}) to binary neutron star mergers, both in the cases of current surveys, such as CDF-S and {\it XMM-Newton} PO and SLEW, and proposed missions, such as {\it THESEUS}.

\texttt{saprEMo} provides theoretical predictions allowing us:
\begin{itemize}
\item {\it{to compare predictions with actual data.}}
E.g. we proved that some signals consistent with the model could already be detected in present surveys of data such as {\it XMM-Newton} PO and SLEW;
\item {\it{to test potential associations.}} E.g. we proved that the new transient found by \citealt{bauer2017new} is marginally consistent with the model;
\item {\it{to assess the effectiveness of proposed mission concepts for a specific type of signal. }}
We illustrate the utility of \texttt{saprEMo} for evaluating proposed missions with a case study of {\it THESEUS}.
We demonstrate that, with few years of operation, the large FoV of {\it THESEUS/SXI} could allow for the detection of up to thousands of SC16-like signals, enabling considerable constrains on both the BNS merger rate and the emission models;
\item {\it{and to compare different observational strategies.}}
\texttt{saprEMo} can be used to determine advantages and disadvantages compared to a particular emission, of adopting different observational strategies.
With the case of {\it THESEUS}, we indeed demonstrate that \texttt{saprEMo} can compare observations characterised by different values of typical exposure time $\left<T_{obs}\right>$ and point out the main properties of the relative detections.
In general, given a total observing time $T$, different observational strategies can be applied; increasing the exposure time to increase the sensitivity or decreasing the exposure time to enlarge the sky coverage.
The effect of adopting different exposures depends on several parameters, including both source and instrumental properties (such as rate, luminosity, flux limit dependence on exposure time, etc).
In this paper, we specifically prove that, given {\it THESEUS/SXI} sensitivity as a function of exposure time, 
10 observations of disjoint sky areas lasting 1~ks would en-captured more SC16f-like transients than an extended single pointing of 10~ks.
\end{itemize}
In general \texttt{saprEMo} allows us to test both survey and astrophysical properties. This study has mainly focused on the former, exploring the impact 
of different trade offs among such properties (including exposure time, sky localisation, and spectral sensitivity), assuming a single light curve model from SC16.
However \texttt{saprEMo} can also
test (and be used for inference on) astrophysical quantities such as emission duration, peak luminosity and spectra.\\
Once the design sensitivity of Advanced gravitational-wave interferometers is achieved, GW detections of EM bright sources, such as GW170817, will occur more and more often and very likely at lower SNRs. 
In this context of multi-messenger astronomy, 
\texttt{saprEMo} can be used to
optimize the analysis 
by identifying specific emission model. 
We conclude remarking that the flexibility of the implemented methodology allows considerations of emission model spanning the whole electromagnetic spectrum (e.g. kilonovae models can also be tested).
Moreover our analysis include no priors on nature of EM sources, so that it can be applied to a wide range of astrophysical phenomena.
With its analysis dedicated to treat high redshift effects, \texttt{saprEMo} particularly suits studies on emission of cosmological origin. 

%%%%%%%%%%%%%%%%%%%%%%%%%%%%%%%%%%%%%%%%%%%%%%%%
%%%%%%%%%%%%%%%%%%%%%%%%%%%%%%%%%%%%%%%%%%%%%%%%%%%
\section*{Acknowledgements}
%%%%%%%%%%%%%%%%%%%%%%%%%%%%%%%%%%%%%%%%%%%%%%%%%%%
The authors thank L. Amati for his assistance with the case of {\it THESEUS} and for the useful comments.
We thank A. Belfiore, A. De Luca, M.Marelli, D. Salvietti and  A. Tiengo for the help in understanding and interpreting {\it XMM-Newton} data. We thank R. Salvaterra for the useful suggestions and discussions. 
The research leading to these results has received funding from the 
People Programme (Marie Curie Actions) of the European Union's Seventh 
Framework Programme FP7/2007-2013/ (PEOPLE-2013-ITN) under REA grant 
agreement no.~[606176]. This paper reflects only the authors' view and the European Union is not liable for any use that may be made of the information contained therein.
G.S. acknowledges EGO support through a VESF fellowship (EGO-DIR-133-2015).
This research has made use of data obtained from XMMSL2, the Second {\it XMM-Newton} Slew Survey Catalogue, produced by members of the XMM SOC, the EPIC consortium, and using work carried out in the context of the EXTraS project ("Exploring the X-ray Transient and variable Sky", funded from the EU's Seventh Framework Programme under grant agreement no.~[607452]).
This research has made use of data obtained from the 3XMM {\it XMM-Newton}  serendipitous source catalogue compiled by the 10 institutes of the {\it XMM-Newton} Survey Science Centre selected by ESA.

%%%%%%%%%%%%   REFERENCES   %%%%%%%%%%%%%%%%%%%%%%%%
\bibliographystyle{mnras} 
\bibliography{bibliography}

%%%%%%%%%%%%%%%%%%%%%%%%%%%%%%%%%%%%%%%%%%%%%%%%%%%
\begin{appendix} %First online appendix
%%%%%%%%%%%%%%%%%%%%%%%%%%%%%%%%%%%%%%%%%%%%%%%%%%%
\section{saprEMo features}
\label{App:A}
%%%%%%%%%%%%%%%%%%%%%%%%%%%%%%
\subsection{Redshift / K-correction}
To evaluate if the emission is visible in the $g$ band of the instrument, we need to 
calculate the fraction of source light curve which contributes to the flux in the $g$ band at the observer.
Here we denote each energy band $[E^I_{min},E_{max}^I]_g$ with the label $g$; we use $[E^I_{min},E_{max}^I]$ without subscripts for the whole range of operation.\\
We assume a set of light curves $[L_h(t')]_{h=0}^{h_{max}}$, where each element $L_h(t')$ represents the emission in the source frame within a fixed $h$ energy bin $[E'_{min},E'_{max}]_h$.
The redshifted energy bin $h$, associated to the light-curve element $L_h(t')$, might only partially overlap with an instrumental energy bin $g$. 
To consider only the part of the light-curve which contributes to the emission visible in the $g$ energy bin,
we calculate the fraction of the $h$ energy bin that falls into the $g$ band and assume the energy is uniform across its intrinsic spectrum.
Therefore for each numerical step in $z$, we calculate the emission contribution to each $g$ band:
\begin{eqnarray}
L_g (t',z) = \sum_{h= 0}^{h_{max}} L_h(t') w_{hg}(z)
\label{eq:Lg}
\end{eqnarray}
When the redshifted $h$ bin and the $g$ band of the instrument overlap, the $L_h(t')$ emission contributes to the total observable emission in the $g$ band $L_g (t',z)$ with a weight 
defined by the ratio between the amount of overlap and the width of the light curve energy bin:
\begin{eqnarray}
w_{hg}(z)= \begin{cases}
 \alpha\, \text{if}\, \alpha>0\\
 0\, \text{otherwise}
\end{cases}
\end{eqnarray}
where 
\begin{eqnarray}
\alpha = \displaystyle\frac{min\left((1+z)^{-1}E'_{max,h},E^I_{max,g}\right) - max\left((1+z)^{-1}E'_{min,h},E^I_{min,g}\right)}{(1+z)^{-1}[E'_{max,h}-E'_{min,h}]}
\end{eqnarray}
An increased resolution in the energy bins of the emission model will result in more precise estimates.
%%%%%%%%%%%%%%%%%%%%%%%%%%%%%%
\subsection{Absorption}
\label{APP_A2}
%%%%%%%%%%%%%%%%%%%%%%%%%%%%%%
\texttt{saprEMo} can account for both host and Galactic absorptions.
The host-galaxy absorption is included by substituting $L_h(t')$ with $\tilde{L}_h(t') = L_h(t')e^{-n_{\mathrm{H,h}}\sigma_h} $, where $n_{\mathrm{H,h}}$ is a typical value of the effective hydrogen column density and $\sigma_h$ is the average of the absorption cross-section in the $h$ energy band in the source frame. Both of these quantities may depend on the type of the host galaxy.
Similarly, the Milky-Way absorption is accounted adopting $\tilde{L}_g(t',z) = L_g(t',z)e^{-n_{\mathrm{H,MW}}\sigma_g}$. 
We estimate an effective hydrogen column density as a function of the observed sky-locations (Galactic latitudes), adopting the sky-map of HI emission-line brightness $T_b$ released by \citealt{kalberla2005leiden}. 
For each position in the sky, the Galactic column density along the line of sight $n_{H,MW}$ is calculated adopting equation (4) of \citealt{chengalur2013accurate} (valid for negligible total opacity).
These values then have to be averaged along Galactic longitudes $l$, $\left<n_{H,MW}\right>_l$, and finally associated to the relative frequency of observations in the survey to calculate an effective column density $n_H$:
\begin{eqnarray}
n_{H,MW} [cm^{-2}] = \frac{1}{n_{obs}}\sum_{j = 1}^{n_{obs}}\left<n_{H,MW}\right>_{l_j}(b_j,l_j)
\end{eqnarray}
where $b$ is the galactic latitude.

To establish the detectability of the light curves $\tilde{L}_g(t',z)$, we calculate the corresponding fluxes: 
\begin{eqnarray}
F_g (t',z) = \frac{\tilde{L}_g (t',z)}{4\pi D_L^2(z)}
\label{eq:Fg}
\end{eqnarray}
\subsubsection{X-ray absorption model}
Different specific absorption models can be implemented accordingly to the energy range of interest.
In this paper we only consider the effect of X-ray absorption at the observer. 
In \citealt{willingale2013calibration},
the authors investigate hundreds of GRB afterglows detected by Swift to model the effective total Galactic column density in X-ray $n_{H_{tot},MW}$.
Atomic and molecular hydrogen represents the 
dominant components of $n_{H_{tot},MW}\approx n_{H,MW} + 2~n_{H_2,MW}$.
To estimate the molecular hydrogen component from the atomic contribution, we adopt the model proposed by 
\citealt{willingale2013calibration}:
\begin{eqnarray}
n_{H_2,MW} = n_{H_2max}\left[1-\exp\left(\frac{n_{H,MW}}{n_c}\right)\right]^\alpha
\end{eqnarray}
where $n_{H_2max} = 7.5 \times 10^{20}~\text{molecules/cm}^{2}$, $n_c = 2.37\times 10^{21}~\text{atoms/cm}^{2}$ and $\alpha = 2$. 
We apply this effective total Galactic column density $n_{H_{tot},MW}$ to both {\it XMM-Newton} surveys, PO and SLEW.

The effective cross section of the interstellar medium, for each of the energy bands in the range $0.03-10$ keV, is analytically estimated as a function of energy $E$, following
\citealt{morrison1983interstellar}.
%%%%%%%%%%%%%%%%%%%%%%%%%%%%%%%%%%%%%%%%%%%%%%%%%%%
%%%%%%%%%%%%%%%%%%%%%%%%%%%%%%
\subsection{Detailed integrations over cosmical scales}
\label{APP_A3}
%%%%%%%%%%%%%%%%%%%%%%%%%%%%%%

To estimate the peak contribution, we need to account for the redshift dependence of all the quantities involved in the cosmic integration of equation \ref{eq:peaks}.
At each step in $z$, the portion of light curve still present (after the redshift) in each of the instrumental energy band is calculated, the absorption is applied and the new maximum redshift $z_{max,g}$ is computed.
The contribution of the correspondent step in redshift $z$ is considered if at least on one energy band $g$, $z_{max,g}>z$.

Similarly, we compute the number of tails with:
\begin{eqnarray}
\begin{split}
N_t = \displaystyle \varepsilon~ \tilde{n}_{obs}\frac{FoV}{4\pi}\int_{0}^{\overline{z_{max}}} R_V(z)\frac{dVc}{dz}LCS\,dz
\label{eq:tails_rev}
\end{split}
\end{eqnarray}
where $\overline{z_{max}}$ is the maximum redshift which contributes to peak count and $LCS(z) = \bigcup_g [dt'_{\mathrm{vis}}(z)]_g$ represents the union of source time-intervals of the light curve $dt'_{\mathrm{vis,i}}$ which are visible at $z$ in 
at least one $g$ band, i.e.
\begin{eqnarray}
\forall g
\quad dt'_{\mathrm{vis,i}} = t'_{i+1} - t'_{i} \mid [F_g(t'_i,z)+F_g(t'_{i+1},z)]>2~F_{lim,g}
\label{eq:dt_vis}
\end{eqnarray}
This contribution does not include peaks, since the peak duration can be consider infinitesimal. The time dependence for tail calculation is only set by the signal duration, while is completely unaffected by the survey exposures, which instead determine the peak contribution.
%%%%%%%%%%%%%%%%%%%%%%%%%%%%%%%%%%%%%%%%%%%%%%%%%%%
%%%%%%%%%%%%%%%%%%%%%%%%%%%%%%%%%%%%%%%%%%%%%%%%%%%
\section{Survey properties}
\label{App:Survey_Prop}
\subsection{XMM-Newton parameters}
%%%%%%%%%%%%%%%%%%%%%%%%%%%%%%%%%%%%%%%%%%%%%%%%%%%
We report the parameters adopted to apply \texttt{saprEMo} to {\it XMM-Newton} surveys: PO and SLEW in tables \ref{tab:XMM_PO_par} , \ref{tab:XMM_PO_sens}, \ref{tab:XMM_SLEW_par} and \ref{tab:XMM_SLEW_sens}. 
The data used to characterise PO and SLEW are collected in source catalogs. 
From there, relative papers \citep{3XMM_Pref,SLEW_Pref} and webpages \citep{3XMM-DR7ref,SLEW_Pref2,SLEW_Pref3,SLEWpropWEB}, we extract the general properties necessary to apply \texttt{saprEMo}. In the case of PO and SLEW we estimate the impact of absorption from the locations of the sources collected in the catalogs  \citealt{3XMM_Pref} \footnote{We adopt clean data, requiring: $\mathrm{CLEAN = OBS\_CLASS<3}$.
Data concerning average and standard deviation of observation durations are based on 
total band ($Exp\_Map\_B8$).
Locations adopted for the absorption model are inferred from source locations.}.
Since no flux limits have been quoted for PO, for that case we use as a proxy the medians of the fluxes available from the catalog.
\label{Appendix:XMM-Newton}
%%%%%%%%%%%%%%%%%%%%%%%%%%%%%%%%%%%%%%%%%%%%%%%%%%%
\subsection{CHANDRA}
\label{Appendix:CHANDRA}
%%%%%%%%%%%%%%%%%%%%%%%%%%%%%%%%%%%%%%%%%%%%%%%%%%%
To apply \texttt{saprEMo} to the CDF-S we use data from  \citealt{CHANDFS7MS,lehmer2005extended} and  \citealt{luo2016chandra}. As quoted in the same references, we adopted $n_{H,MW} = 8.8\times 10^{19}~\mathrm{cm^{-2}}$. 

The adopted survey properties are reported in the tables \ref{tab:CHANDRA_par} and \ref{tab:CHANDRA_sens}.
Tables \ref{tab:CHANDRA_par} and \ref{tab:CHANDRA_sens} report conservative values extracted from figures 2, 28 and 29 of \citealt{luo2016chandra}, approximating the main characteristic a region of the sky 
described by roughly homogeneous properties \footnote{To account for homogeneity, we decide to only consider a 285~arcmin$^2$ region, out of the 484.2~arcmin$^2$ of the entire survey, which corresponds to the ACIS FoV and roughly to the region observed by at least 6 million seconds (see figure 2 of \citealt{luo2016chandra}). A similar sky area is also characterised by flux limits in the total energy band $<2\times 10^{-16}~\mathrm{erg~ cm^{-2} s^{-1}}$.}.
More details on the set of observations are available at \citealt{CHANDFS7MS}.
%%%%%%%%%%%%%%%%%%%%%%%%%%%%%%%%%%%%%%%%%%%%%%%%%%%
\subsection{{\it THESEUS}}
\label{appedix:THESEUS}
%%%%%%%%%%%%%%%%%%%%%%%%%%%%%%%%%%%%%%%%%%%%%%%%%%%
The data concerning {\it THESEUS} have been extrapolated from \citealt{Amati2017}.
We analyse 10~ks of exposure collected with two different strategies:
\begin{description}
\item $\mathbf{a}$) 1 single exposure of 10~ks;
\item $\mathbf{b}$) 10 distinct exposures of 1~ks each.
\end{description}
We report in tables \ref{tab:THESEUS_par}, \ref{tab:THESEUS_sens} the properties adopted for the results presented in this paper.
%%%%%%%%%% XMM NEWTON
\begin{table}
\centering
\begin{tabular}{|c|c|}
{\bf PARAMETER} & {\bf VALUE} \\
\hline
Minimum energy & $0.2~$keV\\
\hline
Maximum energy& $12~$keV \\
\hline
$\left<T_{obs}\right>$ $\,^{a}$ & $19000~$s\\
\hline
$\sigma_{T_{obs}}$ $\,^{a}$ & $17900~$s\\
\hline
Covered Sky area & $1750~(1032^{b})~$deg$^2$\\
\hline
\end{tabular}
\caption{
{\bf PO:} general characteristics of pointed observations contributing to the {\it XMM-Newton} Serendipitous Source Catalog. The data are part of the 3XMM-DR7 catalogue \citep{3XMM_Pref}; we adopted the fit file $3xmmdr7\_obslist.fits$ available at \citep{TAB2p1_XMM_PO}.
{\it a}:
From clean observations ($\mathrm{OBS\_CLASS}<3$);
{\it b}: Excluding overlaps.
}
\label{tab:XMM_PO_par}
\end{table}
\begin{table}
\centering
\begin{tabular}{|c|c|}
{\bf ENERGY BAND} [keV]& {\bf SENSITITY} $[\mathrm{erg~ cm^{-2}~ s^{-1}}]$\\
\hline
$0.2 -0.5$& $5.8\times10^{-16}$\\
\hline
$0.5 -1.0$& $1.7\times10^{-15}$\\
\hline
$1.0-2.0$ & $2.7\times10^{-15}$\\
\hline
$2.0-4.5$ & $3.8\times10^{-15}$\\
\hline
$4.5-12.0$ & $6.6\times10^{-15}$\\
\hline
\end{tabular}
\caption{
{\bf PO:} spectral bands of pointed observations contributing to the {\it XMM-Newton} Serendipitous Source Catalog. Medians in each band from catalog as reported in website \citep{heasarc3XMMDR7}, catalog available at \citep{3XMM-DR7cat} 
catalog: $3XMM\_DR7$ cleaned with the same criteria used for calculating average and variance of exposure times.
}
\label{tab:XMM_PO_sens}
\end{table}
\begin{table}
\centering
\begin{tabular}{|c|c|}
{\bf PARAMETER} & {\bf VALUE }\\
\hline
Minimum energy & $0.2~$keV\\
\hline
Maximum energy& $12~$keV \\
\hline
$\left<T_{obs}\right>$ $\,^{a}$ & $6.9~$s\\
\hline
$\sigma_{T_{obs}}$ $\,^{a}$ & $2.4~$s\\
\hline
Covered Sky area$^b$ & $84\%$ of the sky\\
\hline
\end{tabular}
\caption{
{\bf SLEW:} general characteristics of slew data contributing to the {\it XMM-Newton} Slew Survey Catalogue.
{\it a}: from clean observations according to $xmmsl2\_clean.fits$ file at  \citep{SLEWref};
{\it b}: percentage when overlaps are excluded from \citep{SLEWref}.
}
\label{tab:XMM_SLEW_par}
\end{table}

\begin{table}
\centering
\begin{tabular}{|c|c|}
{\bf ENERGY BAND} [keV]& {\bf SENSITITY} [$\mathrm{erg~ cm^{-2}~ s^{-1}}$]\\
\hline
$0.2 -2.0$& $0.57\times10^{-12}$\\
\hline
$2.0 -12.0$& $3.7\times10^{-12}$\\
\hline
\end{tabular}
\caption{
{\bf SLEW:} spectral bands of slew data contributing to the {\it XMM-Newton}  Slew Survey Catalogue. 
Flux limits from \citep{SLEWpropWEB}.
The energy range in SLEW catalog is divided just in 2 bands.
}
\label{tab:XMM_SLEW_sens}
\end{table}
%%%%%%%%%%%%%%%%%%%%%%%%%%%%%%%%%%%%%%%%%%%%%%%%
%%%%%%%%%%%%%%%%%%%%%%%% CHANDRA %%%%%%%%%%%%%%%%%%%%%%%%
\begin{table}
\centering
\begin{tabular}{|c|c|}
{\bf PARAMETER} & {\bf VALUE} \\
\hline
Minimum energy & $0.5~$keV\\
\hline
Maximum energy& $7~$keV \\
\hline
FoV & $285~$arcmin$^2$\\
\hline
{\color{red}} $\left<T_{obs}\right>^a$ & $\sim 6\times 10^6~$[s]\\
\hline
$n_{obs}$ & 1\\
\hline
\end{tabular}
\caption{
{\bf CDF-S:} general properties of CDF-S (roughly homogeneous region) from \citep{lehmer2005extended}, and \citep{luo2016chandra}.
$^a$: the maximum cleaned exposure is $6.727\times 10^6~$s.
}
\label{tab:CHANDRA_par}
\end{table}

\begin{table}
\centering
\begin{tabular}{|c|c|}
\hline
{\bf ENERGY BANDS} [keV]& {\bf SENSITIVITY }[$\mathrm{erg~ cm^{-2}~ s^{-1}}$]\\
\hline
$0.5 -2.0$& $\sim 6\times10^{-17}$\\
\hline
$2.0 -7.0$& $\sim 4 \times10^{-16}$\\
\hline
\end{tabular}
\caption{
{\bf CDF-S:} conservative estimates of flux limits from figures 28 and 29 of \citet{luo2016chandra} (at $\sim$50\% completeness).
}
\label{tab:CHANDRA_sens}
\end{table}

%%%%%%%%%%%%%%%%%%%%%%%%%%%%%%%%%%%%%%%%%%%%%%%%
%%%%%%%%%%%%%%%%%%%%%%%% THESEUS %%%%%%%%%%%%%%%%%%%%%%%%
\begin{table}
\centering
\begin{tabular}{|c|c|}
{\bf PARAMETER} & {\bf VALUE} \\
\hline
Minimum energy & $0.3~$keV\\
\hline
Maximum energy& $6~$keV \\
\hline
FoV& $110\times30~$deg$^2$\\
\hline
\end{tabular}
\caption{{\bf {\it THESEUS}:} general properties of the Soft X-ray Imager (SXI) from \citep{Amati2017}.}
\label{tab:THESEUS_par}
\end{table}

\begin{table}
\centering
\begin{tabular}{|c|c|c|c|}
&$\left<\mathbf{T_{obs}}\right>$ [$s$]& $\mathbf{n_{obs}}$ & {\bf SXI SENSITIVITY} [$\mathrm{erg~ cm^{-2}~ s^{-1}}$]\\
\hline
{\bf Case} $\mathbf{a.}$ &$10^4$& 1& $7.82\times10^{-12}$  $(1.93\times10^{-11})$\\
\hline
{\bf Case } $\mathbf{b.}$ &$10^3$& 10& $3.20\times10^{-11}$ $(7.87\times10^{-11})$\\
\hline
\end{tabular}
\caption{{\bf {\it THESEUS}:} properties of the two considered scenarios for a total of 10ks observations. Sensitivities extrapolated from figure 4 of \citet{Amati2017} assuming column density typical of regions outside (inside) the Galactic plane.}
\label{tab:THESEUS_sens}
\end{table}
\end{appendix}
\label{lastpage}
\end{document}